\documentclass{article}
\usepackage{a4wide}
\usepackage{amsmath,caption}
\usepackage[latin1]{inputenc}
\usepackage{epsfig,array,xspace}

\newcommand{\texthack}[1]{\text{#1}}

\newcommand{\sref}[1]{section \ref{#1}}

\newcommand{\bi}[1]{\textbf{\textit{#1}}}
\newcommand{\latticesites}{L} 
\newcommand{\latticeconst}{\delta}
\newcommand{\vesselset}{V}

\newcommand{\vesselnodeset}{N}

\newcommand{\tumset}{T}

\newcommand{\tumsetUO}{T_{uo}}
\newcommand{\tumradius}{x_{T}}
\newcommand{\vecr}{ \bi{x} }
\newcommand{\vecrprime}{ \bi{x}^\prime }
\newcommand{\foxy}{c_{o}}
\newcommand{\foxyb}{c_{o}^{(B)}}
\newcommand{\oxyq}{\alpha}
\newcommand{\oxyqN}{\alpha^{(0)}}
\newcommand{\foxyk}{\kappa}
\newcommand{\foxyktiss}{\kappa^{(N mm  )}}
\newcommand{\foxyktum}{\kappa^{(T)}}

\newcommand{\fgf}{c_{g}}

\newcommand{\paramTimesprmax}{s^{(max)}}
\newcommand{\paramSproutParentInTumor}{t_{EC}^{(switch)}}
\newcommand{\paramTimeSprout}{t_{EC}^{(sprout)}}
\newcommand{\paramTimeDil}{t_{EC}^{(dil)}}
\newcommand{\paramTimeTcProl}{t_{TC}^{(prol)}}
\newcommand{\paramTimeUO}{t_{TC}^{(uo)}}

\newcommand{\paramFCrit}{f^{(c)}}
\newcommand{\paramFCritN}{f^{(c,max)}}
\newcommand{\paramFCritDelta}{\delta^{(c)}}

\newcommand{\paramRMax}{r^{(max)}}
\newcommand{\paramRInit}{r^{(init)}}
\newcommand{\paramProbColl}{p^{(c)}}
\newcommand{\paramProbCollN}{p^{(c,max)}}
\newcommand{\paramProbCollRing}{p^{(c)}_{ring}}
\newcommand{\paramProbCollFull}{p^{(c)}_{full}}
\newcommand{\paramTumInit}{|T(t=0)|}
\newcommand{\paramRGf}{R^{(g)}}
\newcommand{\paramLenSprout}{l^{(spr)}}
\newcommand{\tcDeathOxy}{\theta_o^{(death)}}
\newcommand{\tcProlOxy}{\theta_o^{(prol)}}
\newcommand{\vessProlGf}{\theta_g^{(prol)}}

\newcommand{\paramPressBoundary}{P^{(b)}}

\newcommand{\paramDrugInit}{C^{(init)}}
\newcommand{\idWall}{w}
\newcommand{\paramWallCrit}{\idWall^{(c)}}
\newcommand{\paramWallDecreaseRate}{\Delta\idWall}
\newcommand{\paramMaxCapiAddRad}{r^{(capi)}}
\newcommand{\latticeconstgen}{\delta^{(gen)}}
\newcommand{\timet}{t}
\newcommand{\timeis}[1]{$\timet=$#1h}
\newcommand{\deltat}{\Delta t}

\newcommand{\micron}{$\mu$m\xspace}

\newcommand{\figinline}[1]{Fig.#1}
\newcommand{\meanval}[1]{\langle #1\rangle}
\newcommand{\Otwo}{$\text{O}_\text{2}\,$}

\begin{document}

\begin{titlepage}
{\baselineskip26pt
\centering {\bf\Large 
Hot spot formation in tumor vasculature during tumor growth 
in an arterio-venous-network environment}

\vspace*{2cm}

{\large M.Welter$^\text{a}$, K.Bartha$^\text{b}$, H.Rieger$^{\text{a,}*}$}\\
\vspace*{0.5cm}

\normalsize
$^\text{a}$\textit{Theoretische Physik, Saarland University, 
 PF 151150, 66041 Saarbrücken, Germany}\\
$^\text{b}$\textit{Department of Medical Biochemistry, 
 Semmelweis University, Budapest, Hungary}\\
\vspace*{2cm}	
}

\normalsize
\flushleft{\textbf{Abstract}}\\
\vspace*{1em}
Hot spots in tumors are regions of high vascular density in the center
of the tumor and their analysis is an important diagnostic tool in
cancer treatment. We present a model for vascular remodeling in
tumors predicting that the formation of hot spots correlates with
local inhomogeneities of the original arterio-venous vasculature of
the healthy tissue. Probable locations for hot spots in the late
stages of the tumor are locations of increased blood pressure gradients.
The developing tumor vasculature is non-hierarchical but still 
complex displaying algebraically decaying density distributions.
\vspace*{1em}

\textit{Keywords:} Angiogensis; tumor growth; vascular networks; 
cancer; blood flow.
\vfill

*Corresponding author Tel.: +49\,681\,302\,3969; fax: +49\,681\,302\,4899.\\
\textit{E-mail address:} h.rieger@mx.uni-saarland.de (H.Rieger)\\
\end{titlepage}

\cleardoublepage

\section {Introduction}\label{SEC:BIOINTRO}

Tumors that grow inside vascularized tissues remodel their vascular
environment in a characteristic way. Tumor cells secret growth
factors, the most prominent being vascular endothelial growth factor
(VEGF) that diffuse into the surrounding tissue. This initiates a
vascularization program that is denoted as tumor-induced angiogenesis
(Carmeliet and Jain, 2000). The lumen of blood vessels is surrounded
by endothelial cells (ECs), which express VEGF receptors that
stimulate EC migration and proliferation upon binding to
VEGF. Existing blood vessels generate sprouts that migrate through the
extracellular matrix, along a chemotactic gradient if present or
guided by receptors like those from the Eph family and their ephrin
ligands (Kullander and Klein, 2002; Carmeliet and Tessier-Lavigne,
2005) or just randomly in the absence of any guidance clue. As a
result new blood vessels are formed in the peritumoral region, where
the microvascular density (MVD) is increased. This improves the oxygen
and nutrient supply of the periphery and leads to proliferation
and expansion of the tumor.

Inside the tumor the formation of new vessels deceases in spite of an
abundance of growth factors and the vascularization program is
switched from sprouting angiogenesis to circumferential vessel
growth. Experiments demonstrate that certain guidance molecules like
EphB4 expressed by tumor ECs act as a negative regulator of blood
vessel branching and vascular network formation (Erber et al., 2006).
In addition to a drastic dilation of the blood vessels in the tumor
center a massive vessel regression inside the tumor is observed (Holash
et al, 1999; Maisonpierre et al., 1999) leading to a much lower MVD in
the tumor center. Regression of blood vessels inside the tumor can be
evoked by the presence of angiogenic agonists like angiopoetin (Holash
1999a, 1999b) or by mechanical solid stress in conjunction with a
leaky and disorganized vessel structure leading to vessel collapse.

This scenario deviates from the prevailing picture that most tumors
and metastases originate as small avascular masses that belatedly
induce the development of new blood vessels once they grow to a few
millimeters in size (Folkman, 1971; 1990). Instead tumors rapidly
coopt existing host vessels to form an initially well-vascularized
tumor mass, which in later stages attains its characteristic
morphology, which is characterized by (Paku, 1998; Holash 1999a,
1999b; Döme et al., 2002): The tumor is compartmentalized into a
perimeter region with MVD much higher than in the normal tissue, a
well vascularized tumor periphery, and a tumor center with a very low
MVD. In the tumor center necrotic regions are threaded by a few very
thick vessels surrounded by 100-200 $\mu$m thick cuffs of viable tumor
cells. This is expected to be the general course of development of a
tumor vessel network that grow in a vascularized tissue.

Recently two of us introduced a mathematical model (Bartha and Rieger,
2006) predicting the dynamics of vascular network formation based on
the mechanisms identified experimentally and described in part
above. Its basic ingredients are a dynamic network, in which links,
representing vessels, can be added or removed or altered (in diameter
and other parameters), a cellular automaton representing the growing
tumor, and two concentration fields mediating the interaction between
the network and the tumor cells. The links of the network, modeled as
pipes carrying an ideal pipe flow, constitute the sources of the
oxygen or nutrient concentration field, and the tumor cells are
sources of the growth factor concentration field. Generation and
removal of vessels or tumor cells are stochastic processes with
transition probabilities that depend on local concentrations of oxygen
and growth factor, blood flow and duration of under-oxygenation. The
model displays the characteristic features of tumor vasculature
described above and it is robust against parameter variations and
various refinements (Welter et al., 2007) as well as the choice of
space dimensionality (Lee et al., 2006).

Here, the global characteristics of the emerging morphology of the
tumor vasculature, like the radial dependencies of MVD, tumor density,
blood flow etc., were independent of the details of the original
vasculature. However, local characteristics, like the formation of
highly vascularized regions in the tumor center, the so called ``hot
spots'' (Weidner, 1995; Belien et al., 1999), could not be analyzed
since the original vessel network was assumed to form a regular
lattice and the original blood flow imprinted via boundary conditions
was directed along the diagonal of the underlying lattice. As a result,
the surviving thick vessels threading the fully developed tumor where
also running along this diagonal and these represented the only hot
spot, always occurring in the same location of the tumor
network. Since the analysis of hot spots are an important diagnostic
tool in cancer treatment (Weidner, 1995; Belien at al., 1999; Pahernik
et al., 2001) it is highly desirable to understand the
mechanisms determining their development in growing tumors.

The main hypothesis we intend to push forward in this work is that hot
spot formation in tumors is connected to local characteristics of the
original vascular network. For this reason it appears mandatory to
start with a realistic initial vasculature, reflecting the typical
arterio-venous organization of the blood vessel network in healthy
tissues. The construction of such a network consisting of two
interdigitating hierarchical trees, an arterial and a venous tree,
under the constraint that the lowest level, the capillaries, provide a
homogeneous supply of oxygen in the tissue, is a highly non-trivial
task already in a two-dimensional set-up. A mathematical model for the
stochastic generation of an arterio-venous network was developed in
(Gödde and Kurz, 2001) and a modified version of it is used as the
starting configuration of the vessel network in our model. This is
based on the ideas developed in our previous work (Welter et al, 2007)
for a regular starting network and adapted to the presence of arteries
and veins with varying vessel radius and blood flow parameters. The
purpose of this work is to describe the transformation of the initial
arterio-venous vasculature into a non-hierachical tumor vasculature
with that displays still complex properties characterized by
algebraically decaying density distributions. In particular it leads
to the formation of hot spots and the most probable location of the
regions with highest vascular density can be predicted from an
analysis of pressure gradients in the original network.


The paper is organized as follows: In the next section the model is
defined, including the method to construct an initial arterio-venous
network. Section 3 presents the results for global properties of the
tumor vasculature for a choice of parameters that is guided by
experimental data for melanoma. These results include a discussion of
the emerging morphologies; radial distributions of vessel density,
vessel radius, tumor density, flow rates, shear forces etc., vessel
statistics, parameter dependencies and drug flow. In section 4 we
analyze the spatial inhomogeneities and the hot spot formation of the
tumor vasculature. Section 5 concludes the paper with a discussion.

\section{Mathematical Model}

Our mathematical model contains discrete elements and continuous
fields. The tumor is represented by a set of occupied sites on a
lattice and the vascular system by a network of connected tubes which
occupy lattice bonds. The continuous part involves a \Otwo field as
nutrient supply and a growth factor (GF) field for angiogenic
signaling. The temporal evolution is governed by stochastic
processes according to which the automaton state is successively
updated.  An exhaustive description is given in our previous
publication (Welter et~al., 2007), but since the introduction of an
arterio-venous network and some refinements have lead to a few changes
we give a comprehensive overview here.

\subsection{Definition of the system state}

As illustrated in \figinline{1}, we study a quasi two-dimensional
case, where the arrangement of vessels and tumor cells (TCs) is
subject to a regular triangular lattice with lattice-constant
$\latticeconst$ and size $l \times l = n$. Each lattice site can be
occupied each with a single tumor cell. We define $\tumset$ as the set
of sites currently occupied by viable TCs. Non-occupied sites
represent normal tissue, which we assume to be homogeneous although
this is not the case in reality. Likewise lattice bonds can be
occupied with vessels.  The vasculature is modeled as network of
connected ideal tubes.  We can formally describe its topology by a
graph $G =(\vesselnodeset,\vesselset)$, where the edges in
$\vesselset=\left\{ (i,j),\, i,j \in \vesselnodeset \right\}$
represent vessel segments which we also denote as vessels. The
elements in $\vesselnodeset$ refer to attached junction nodes.  Note
that vessels run parallel to the lattice axes and that node locations
coincide with lattice sites.  We further refer to various properties
via sub-scripts, for example the site location of a node $i$ is given
by $\vecr_i$, while $q_a$,$r_a$,$l_a$ denote flow-rate, radius and
length of some vessel $a$.

Blood flow is an important part of our modeling. We assume
steady-state laminar flow through the network. Given the
pressure gradient $\Delta p = (p_i-p_j)/l$ between the ends of a
vessel $(i,j)$, Hagen-Poiseuille's Law determines the flow rate $q
\propto r^4\Delta p / \eta$ and wall shear-stress $f_a \propto r
\Delta p$.  To account for the complex non-Newtonian fluid behavior of
blood it is common to introduce an effective viscosity $\eta =
\eta_{plasma}\,\eta_{rel}(H,r)$ depending on local vessel radius $r$
and hematocrit $H$. We use the empirical formula derived by Pries
et~al. (1994), where the precise definition can be found. For
simplicity we neglect the phase separation effect and set the
hematocrit to $0.45$, the average in the human body. To find the nodal
pressures we solve the system of linear equations which arises from
the application of mass-conservation at junctions.
Appropriate boundary conditions must be present. Our arterio-venous
trees require them only at the ends of the major supply-and drain
vessels. We have chosen to set the respective nodal pressures to fixed
values $\paramPressBoundary(r)$ based on the vessel radius.

Vasculature and tumor interact via the growth factor field $\fgf$ and
the \Otwo field $\foxy$. The latter is given by the reaction diffusion
equation $\Delta \foxy - \foxyk \foxy + \oxyq(\foxyb-\foxy) = 0$,
where sources are present at vessel locations and sinks are
distributed everywhere.  We solve for steady states, because the
relaxation time is of the order of seconds while configuration changes
at the cellular level take hours.  The coefficients have the following
meaning: $\foxyk$ is a tissue dependent consumption rate, $\foxyb$ the
blood oxygen level and $\oxyq$ the source strength coefficient
incorporating local vascular area-fraction and wall
permeability. Thereby a number of assumptions were made: (i) \Otwo
uptake is supply-limited so that a linear approximation to the
low-P\Otwo regime of a Michaelis-Menten relationship is
reasonable. (ii) Blood-P\Otwo is constant throughout the
vasculature. (iii) Vessels contribute equally to \Otwo supply,
i.e. $\oxyq$ is a constant $\oxyqN>0$ at occupied sites, ignoring
differences in wall thickness, surface area, etc.  Note that the
presence of the sink-term $\foxyk\foxy$ ($\foxyk>0$) would lead to
exponentially decaying distributions around point sources. Indeed,
experiments show an approximately linear P\Otwo decay from the outer
vessel wall ranging ca. 150 \micron into the tissue.

In analogy, we assume that due to binding and degradation GF has a
limited range which is explicitly given by $\paramRGf$. Let us denote
$\tumsetUO$ as the set of underoxygenized sites among $\vecr \in \tumset$.
These are characterized by $\foxy(\vecr) < \tcProlOxy$ where
$\tcProlOxy$ is a threshold parameter.  Under the assumption that TCs
produce GF at a constant rate the formulation of a diffusion-reaction
equation leads to a simple solution via a Greens function approach
where $\fgf$ is given as $\fgf(\vecr) = \sum_{\vecrprime \in
\tumsetUO} \tilde{g}(\vecr-\vecrprime)$.  For simplicity, convenience
and efficiency, we use a linearly decaying ``Greens function'':
$\tilde{g}(r) \propto \max\left[ 0,\ 1-r/\paramRGf \right]$.

\subsection{Definition of the dynamical processes}
\label{SEC:DYNAMICAL_PROCESSES}

Tumor and vascular network dynamics are governed by the  
stochastic processes described below. In practice, when we analyze the model by
Monte-Carlo simulations, we iteratively update
the system by a fixed time step $\deltat=1$h. Per step, a sweep
is done for each process. After that the configuration has changed 
and therefore \Otwo and GF fields are recomputed. This is reasonable
if $\delta$ is sufficiently smaller than the rates at which changes
are introduced.

\newcommand{\tumfigref}[1]{(\textit{\figinline{2}#1)}}

\begin{description}
\item[TC Proliferation:] New TCs occupy empty neighbor 
sites $\vecr \notin \tumset$ with probability $\Delta
t/\paramTimeTcProl$ if the local oxygen level is high enough
$\foxy(\vecr) > \tcProlOxy$. This resembles the behavior of real 
tumors (Bru et~al., 2003; Drasdo and Höhme, 2005) where proliferation 
is restricted to a small band behind the invasive  
edge due to space and nutrient constraints.
\tumfigref{a}

\item[TC Death:] TCs are removed with probability 
$p_{TC}^{(Death)}=1/2$ if the local \Otwo level is less than
$\tcDeathOxy$ for longer than $\paramTimeUO$. Since TCs adapt to low
oxygen conditions (Iyer et~al., 1998), $\tcDeathOxy = \tcProlOxy/10$
is very small. For simplicity the survival time $\paramTimeUO$ has
been given a fixed value, in spite of strong variations among genotypes
(Yu et~al., 2002). \tumfigref{b}

\item[Angiogenesis:]
The presence of GF stimulates angiogenic sprouting at a vessel
occupied site $\vecr$ in the following way. First it is required that
the local GF concentration $\fgf(\vecr)$ is larger than the threshold
$\vessProlGf$. Evidently no sprouting happens inside the tumor (Holash
et~al., 1999).  Indeed molecular pathways (via EphB4 and its ligand
ephrinB2, expressed in tumor endothelial cells) have been identified
recently which are related to the switch from sprouting to
circumferential growth (Erber et~al., 2006).  Therefore it is also
required that parent vessels have not been inside the tumor for longer
than $\paramSproutParentInTumor$. We characterize a position as
``inside'' if radial distance to the tumor center is smaller than the
extent of the tumor in the respective direction. A vessel is inside
the tumor if the midpoints of at least 50\% of all occupied bond are
inside.  Our vessels carry a type information: they are either artery,
vein or capillary. Vessels created via sprouting as described here are
always of capillary type, while the type of the initial vessels is
given by construction of the arterio-venous trees. For our base case,
we require the parent vessel to be either of venous or capillary type,
automatically including active (capillary-)sprouts. These spouts can 
thus already branch out into sub-sprouts. We exclude arteries because
the thicker endothelium and presence of smooth-muscle cells makes
sprouting less likely. The last requirement is that adjacent
bifurcations must be at least $\paramLenSprout$ \micron away since
endothelial cells close to bifurcations also do not form sprouts.  If
those criteria are fulfilled an initial segment of length
$\latticeconst$ and radius $\paramRInit$ is appended with probability
$\Delta t/\paramTimeSprout$, possibly splitting the parent vessel. At
successive iterations more such segments are added to the tip with
probability $\Delta t/\paramTimeSprout$. On contact, it connects to
other vessels forming a potentially blood perfused loop.  Thereby the
initial sprout direction is given by the steepest GF descent (Gerhardt
et~al., 2003). Furthermore, sprouts cannot extend indefinitely (Nehls
et~al., 1998). Therefore a ``countdown'' of $\paramTimesprmax$ hours
terminates the sprouting behavior upon expiration (see below). The
respective internal timer variable is inherited from the previous
sprouting tip.  Note that tumors grow by co-opting vessels but no
explicit modeling is required for this since TCs proliferate in a
separate layer without mechanical interaction.
\tumfigref{c,d} 

\item[Vessel Collapse:]
Vessel in normal arterio-venous networks have differing amounts of
support structures around the endothelium: the basement membrane,
pericytes and smooth-muscle cells.  In tumors interaction/adhesion
between these components is often disrupted (McDonald and Choyke,
2003). There one finds vessels with fragmented and multi-layered 
basement membranes and detached pericytes. Further the lumen
of such vessels is often completely collapsed and/or the endothelium
is also found in a state of regression.  Therefore, we believe that it
is physiologically sound that the survival time of initially healthy
vessels depends on how well the vessel wall is developed,
i.e. capillaries with just the membrane around the endothelium should
collapse earlier than thicker arterioles.  We realize this by a
``degree-of-maturation'' variable $\idWall$ which is initialized with
data for the wall thickness in normal vessels (Pries et~al., 2005),
depending on vessel radius $r$. Using a rough approximation, we set
$\idWall=2r(0.65-0.2\log(2r))$. Inside the tumor, vessels are
continuously degraded at the rate $\paramWallDecreaseRate$. While
$\idWall$ is larger than the threshold $\paramWallCrit$, a vessel does
not regress, meaning the segment is not removed from the network.
Furthermore, physiological wall shear stress levels can dose
dependently inhibit endothelial cell apoptosis, while long term
reduction can cause vessel regression (Dimmeler and Zeiher, 2000). 
We include this by constraining removals to vessels with shear forces 
$f$ less than the threshold $\paramFCrit$. In
tumors, impaired blood flow and thereby low shear stresses might be
the result of increased solid pressure compressing the vessels.  Also
our previous work has shown that shear force correlated collapses lead
to realistic network morphologies. 
If collapse
inhibiting effects do not apply, segments are removed with probability
$\paramProbColl$.  For a dose dependent effect we set
$\paramProbColl = \paramProbCollFull = \paramProbCollN\,(1.0-f/\paramFCrit)$, 
with the constant parameter $\paramProbCollN$. Note that sprouts are
completely excluded from the above processes. In \sref{SEC:PARAMETERVARI}
we also discuss the effect of restricting collapses to a thin band
behind the invasive edge, using the collapse probability $\paramProbCollRing$
as defined there.
\tumfigref{e}

\item[Vessel Dilation:]
Exposure to growth factors is also associated with unphysiological
vessel dilation, in particular in tumors which is related to the
mentioned switching to circumferential growth inside the tumor (Erber
et~al., 2006). Experiments (Döme et~al., 2002) indicate that vessel
diameters increase continually to an upper limit.  In the model the
dilation process works as follows: Each bond occupied by a non
sprouting vessel $a$ increases the radius $r_a$ by $\Delta r_a$ with
probability $\Delta t/\paramTimeDil$ if $r_a<\paramRMax$, if the
center of the bond is inside the tumor and if the local GF
concentration is larger than $\tcProlOxy$.  The increment $\Delta r_a
=\latticeconst^2/2\pi l_a$ corresponds to the surface area of an
additional EC contributed to the total surface area of the segment. To
account for surface tension, a smoothing effect is generated by
dilating the thinnest vessel at junction bonds.
\tumfigref{f}
\end{description}

\subsection{Arterio-venous tree generation}

\newcommand{\godde}{G\&K}

Vasculature in living tissue exhibits a tree like structure. 
Few thick arteries branch out into arteriolar microvessels. 
Terminal branches are connected to the capillary bed, a dense 
network consisting of thin vessels where most of the exchange 
with the surrounding tissue happens. Further upstream blood 
is collected in venoules which fuse into thick veins. In the 
sense of some optimality measure, the design goal of such 
structure is to provide sufficient amount of nutrients to tissue
,while minimizing the effort to keep the circulatory system 
operating.

Numerous ways to generate vasculature in silico have been proposed in
the literature for example fractal approaches based on L-Systems
(Mandelbrot, 1991) or optimization based geometrical construction
(Schrener and Buxbaum, 1993). Most of which concentrate on building
the supplying arteries only, neglecting explicit modeling of
capillaries and veins.  A deterministic manual construction of a
single configuration would have been unsuitable and difficult with
respect to the desire to obtain the flexibility to create a wide range
of configurations while retaining physiological properties.

Gödde and Kurz, (2001) presented a method to grow and remodel vascular
trees stochastically according to probability functions that depend on
local system properties. The simultaneous construction of arterial and
venous trees makes their approach well suited for generating an
initial vasculature for our tumor simulator, where a complete network
including draining venoules is required.

The general idea is to grow purely random arterial- and veneous
trees on a lattice, under exclusion of already occupied sites. 
Followed by iterative shear-stress guided further growth and regression
at the tree leafs, whereby a well perfused space filling network
evolves.

\subsubsection{Network construction}

The mathematical model for the vasculature is identical to the 
definition in the tumor simulation: the network is identified with 
a graph plus additional geometric and hemodynamic properties, 
embedded in a triangular lattice.

Our initial condition for the growth stage consists of several 
prescribed segment chains for arteries and veins which become the 
major vessels in the upper tree levels. Their length and position 
is drawn from a uniform distribution within reasonable bounds.

The basic structural element for further growth is a tripod consisting 
of three vessel segments arranged in $120^\circ$ angles. These tripods 
are appended successively at randomly chosen tree leafs. Exceptions 
are the initial chains where all included sites are candidates for 
adding elements. Thereby overlap with already occupied sites/bonds is 
forbidden, and so the process continues until there are no more valid 
configurations where tripods could be added. The resulting network consists 
of at least two binary trees with at least one arterial and one venous side.

To compute blood flow
vessel radii must be known. `Murray's Law' (Murray, 1926) 
relates the radius of a parent vessel $a$ to the radi of branches $b$,$c$. 
$a_r^\alpha = b_r^\alpha + c_r^\alpha$, where $\alpha = 2.7$ is a realistic
value as used by Gödde and Kurz.
The radii of the tree leafs are preset to 4 \micron for arteries, and 
5 \micron for veins. Thus all other radii can be computed in a recursive 
algorithm which visits a vessels once its descendants have been processed. 

To get a complete network, capillary segments have to be added. 
We simply loop through the bonds on the lattice and add a connection 
$a$ with $a_r=3.5$ \micron if it connects an arterial tree with a 
venous tree. As additional constraint a connection is not added if more 
than three vessels would meet at a site. Capillaries are not constraint 
to connect leafs nodes exclusively, however all vessels at the respective node 
pair must fulfill $r < \paramMaxCapiAddRad$.

The second stage of construction consists of iterating: radii determination, 
capillary creation, flow rate and shear force computation, capillary removal, 
and remodeling sweep, until the system reaches a steady state.
Our remodeling procedure works as follows: One of ${ growth,\, death,\, idle }$ 
is drawn for each segment according to the respective probabilities $p_g,p_d,1-p_g-p_d$, 
where $p_g+p_d \le 1$. Relating these to shear stress $f$ such that $p_g 
\nearrow f$ and $g_d \searrow f$ leads to expansion in well perfused 
branches while non-perfused branches regress and make room for growth of 
interdigitating patterns. The probability relationship we used here is as follows:
\begin{eqnarray*}
	\tilde{f} & \gets \frac{ \ln f - \min_a \left\{ \ln f_a \right\}}
	{\max_a \left\{ \ln f_a \right\} - \min_a \left\{ \ln f_a \right\}} \\
	\tilde{r} & \gets 1 - \min \left\{ 1, \frac{r-4\mu m}{10-4\mu m} \right\} \\
	\tilde{p_g} & \gets ( \tilde{f} + p_0 ) \cdot \tilde{r}\\
	\tilde{g_d} & \gets ( 1-\tilde{f}  + p_0 ) \\
	p_g & \gets \tilde{p_g}/( \tilde{p_g} + \tilde{p_d} )\\
	p_d & \gets \tilde{p_g}/( \tilde{p_d} + \tilde{p_d} )\\
\end{eqnarray*}
To prevent unnatural situations where thick vessels have too many capillary 
branches while still obtaining homogeneous MVD there is the radius term $\tilde{r}$ 
which decays from 1 at $r=4$ to 0 at $10$ \micron. The offset $p_0$ is 
added for two reasons: to lower the rate at which un-circulated initial 
vessels regress, and to add fluctuations to reduce deadlocks in high shear 
stress regions. 
Leaf segments labeled as dying are removed. The removal is constraint 
to leafs in order to maintain the binary tree structure. Next, new elements are added 
to growing segments, whereby one of all admissible configurations is randomly
picked. Beside tripods we also allow single segments to be added in order to allow
vascularization of regions where bottlenecks in between major 
vessels would block access.

\subsubsection{Construction results}

In order to study typical tumor-network morphologies 
we must simulate a sufficiently large domain: 12$\times$10 mm. 
To obtain a specific MVD we adjust the tree generators bond-length
$\latticeconstgen$ and set the lattice size accordingly. 
Due to the two-dimensionality of our system we measure MVD
as the fraction of occupied sites.
Furthermore we chose $\latticeconstgen$ to be a multiple of the
tumor lattice bond-length $\latticeconst$ so that the network
can be taken to the tumor model by superimposing the lattices.
Based on the average for human skin, 100 vessels per mm$^2$ cross-section (Döme et~al., 2002)
which implies 10 \micron inter-vessel spacing, 
we set $\latticeconstgen=60$ \micron and $l=200$.

Due to the hierarchical design of real vasculature, blood-pressure
correlates well with vessel radius. Plotted, it exhibits a sigmoidal
shape (Pries et~al., 1995). 
In order to prevent physiologically inconsistent pressure gradients 
across our networks, we adapt the boundary pressure by an approximate formula.
($p(r)=0.133(18 + 72/(1+\exp((r+21\mu m)/16\mu m)))$ kPa; $r$ is negated 
if it is the radius of an arterial segment). 

As illustrated in \figinline{3}, we have chosen four basic 
layouts. The inlet schematics indicate bounds for the positions 
and lengths of the initial chains. These are drawn from uniform 
distributions within a prescribed range, whereas their types
(arterial or venous) are unalterable. The initial conditions were 
deliberately chosen, motivated by the observation that vasculatures
exhibit relatively thick straight vessels down to a certain level
in the hierarchy as can be seen in photos from the CAM (Mironov et~al., 1998),
or in recent 3d vascular imaging experiments (Cassot et~al., 2006).

Quantitatively we have compared our data with scatter plots of hemodynamic 
properties shown in (Gödde and Kurz, 2001). They are near identical to 
the original and therefore not shown directly. Instead we refer to 
\figinline{10} where tumor vessels are also included.

\subsection{Parameters}
\label{SEC:PARAMETERS}

In what follows we analyze the model with parameters guided by
experimental data (Döme et~al., 2002) for human malignant melanomas.

The lattice bond length $\latticeconst$ is set to 10 \micron, the
typical diameter of TCs and ECs. The initial tumor is an eden-grown
cluster of 1000 cells and 300 \micron in diameter.  The initial
vasculature has been discussed. Growth and death rate parameters as
well as oxygen and growth factor diffusion parameters have been
discussed in detail in (Welter et~al., 2007).

To recapitulate, we set the TC proliferation time $\paramTimeTcProl$
to 10 h, sprout initiation/extension time $\paramTimeSprout$ to 5 h, the time
until regression of loose sprouts $\paramTimesprmax$ to 100 h (Nehls
et~al. 1998), the minimum distance between bifurcations
$\paramLenSprout$ to 20 \micron, the initial sprout radius
$\paramRInit$ to 4 \micron, the time until sprouting switches to
circumferential growth $\paramSproutParentInTumor$ to 24 h, EC
proliferation time for circumferential growth $\paramTimeDil$ to 40 h
and the maximum radius $\paramRMax$ to 25 \micron.  For the continuous
fields we set the \Otwo source coefficient $\oxyqN$ to 0.002, the
consumption coefficient $\foxyk$ to $\foxyktiss=(90$ \micron$)^{-2}$
in normal tissue and to $\foxyktum = 4\foxyktiss$ in tumor tissue to
get halve the diffusion distance. Here the blood oxygen level $\foxyb$
can be chosen arbitrarily because it appears as global scaling factor
in the solution. The model however can be adjusted accordingly by
scaling the threshold parameters. 
Therefore we set it as previously
to the hematocrit value $\foxyb=0.45$ (which is constant here).  We
set the threshold for TC proliferation $\tcProlOxy$ to $0.3 \approx
0.9\meanval{\foxy}$ and threshold for extreme under-oxygenation
$\tcDeathOxy$ much smaller to $\tcProlOxy/10$.  The TC survival time
under hypoxia $\paramTimeUO$ is set to 100 h.  The growth
factor radius $\paramRGf$ is set to 200 \micron, and the vessel
proliferation threshold $\vessProlGf$ is set very low to $10^{-4}$ so
that all vessels within the full GF radius are affected. Parameters
related to vessel regression are as follows. The critical shear force
$\paramFCritN$ is set to $1$ Pa. Note that average shear stress is
$\meanval{f} \approx 10$ Pa.  The collapse probability $\paramProbCollN$ 
is set relatively high to $0.1$.  The dematuration rate
$\paramWallDecreaseRate$ is set to $0.04$ \micron/h, resulting in
regression delays from 93h (4\micron vessels) to 625h (50\micron
vessels). 

\section{Results}

We run 10 Monte-Carla simulations for each of the four initial
vascular setups as explained above and shown in
\figinline{3}. We generated different initial networks for
each run by changing the random number seed for the construction
algorithm.  The vessel density distribution in the vascularized
regions is generally very homogeneous by design since we attempt to
maximize the lattice occupation. Normally the networks fill less than
the entire rectangular domains. We accept this if the support area is
sufficient for tumor growth till the end at \timeis{1200}

\figinline{4} shows the evolution of the model 
using vasculature configuration (a). Not shown is the very first stage
of tumor growth which includes the following.  High tumor \Otwo
consumption leads to decreased \Otwo levels inside the nucleus and
consequently enables vascular remodeling via growth factor
production. Angiogenic sprouting increases the MVD around the tumor,
while after a few hours of prolonged exposure to GFs, central
capillaries show noticeable increased diameters. The first image shows
the tumor at \timeis{200} where a dense capillary plexus has been
created by angiogenic sprouting around the tumor nucleus. Vessel
collapses are not yet occurring there. By \timeis{100} the wall
maturation $\idWall$ decreased sufficiently though to allows first
collapses of the thinnest capillaries with inadequate shear force. We
observe this in other simulation runs.  The process of angiogenesis
and collapse proceeds as the tumor expands. At
\timeis{400} first regions are visible where TCs have died due to hypoxia.
Thick arterioles and venoules don't yet collapse due to
$\idWall>\paramWallCrit$, though they may have low shear force levels.
\figinline{5} shows the final tumor at \timeis{1200} and 
\figinline{6} shows tumors of the other network configurations.
\figinline{7} shows a magnified view on the boundary region 
of the tumor in \figinline{5}.  The tumor masses expand
approximately disc shaped. Because of the simple oxygen distribution
model where all vessels release \Otwo homogeneously, this fulfills our
expectations.  Normally nutrient exchange happens at the thinnest
capillaries. Venoules and arterioles would naturally not contribute
to tissue supply as much. This however is not practical in a two
dimensional model because non-oxygen-releasing vessels would be
impassable barriers to the tumor. Therefore we require that \Otwo
content, and wall-permeability is equal for all vessel.  The resulting
network morphologies show the typical high MVD periphery/peritumoral
region and low density center. A general feature of our model seems to
be the formation of arterio-veneous shunts cause by dilation of
vascular pathways that connect to initially available
arterioles/venoules. Those paths can consist of neo-vasculature as well
as initial capillaries and other vessels. As we expect, most vessels
form isolated straight threads, however regions where vessels are
densely clustered are also observable.

\subsection{Radial distributions}
For comparison \figinline{9} shows quantities averaged within concentric
shells emanating from the tumor center. MVD and TC density are computed as the 
fraction of occupied sites within a shell. Hemodynamic properties are averaged 
over occupied bonds only. Mean values of the microvascular density $MVD$ and 
vessel radius $r$, reminiscent of results from earlier work, agree well with 
experiments (Döme et~al., 2002) as we have discussed for similar data in 
(Bartha and Rieger, 2006).
Döme et~al. measured the MVD and MVR in three
distinct regions: the central region, a $100 \mu m$ wide peripheral
band just behind the invasive edge, a $200 \mu m$ wide peritumoral
region outside the invasive edge. In the central region, they found
$25\%$ MVD of normal tissue, and up to $200\%$ in the peritumoral
region. The vessel perimeter grows linearly from $50 \mu m$ and
assumes a plateau at $200\mu m$ by day 15.

In contrast to results for a regular vascular network (Bartha and
Rieger, 2006; Welter et al., 2007), flow rates and shear force now
show a plateau like the vessel radius which is expected since the flow
boundary condition on the regular network lead to unrealistic star
shaped morphologies, directing all blood flow through the center. The
hierarchical networks here do not introduce such obvious problems.
Not shown is data for the blood-pressure gradient $dp/dl$ which
decreases monotonically toward the center by more than one order of
magnitude. The blood flow rate $q$ is proportional to $r^4 dp/dl$ and
increases on average toward the center beyond a small local minimum at
the invasive edge. Thus the $r^4$ dependency of the radius outweighs
the pressure drop. The shear force dependency is $r dp/dl$, leading to
a decrease respectively.

\subsection{Vessel Statistics}

\figinline{10} shows scatter plots of hemodynamic 
variables against the vessel radius $r$. While providing more
information about their distribution over vessels, such plots are also
common in the literature. Compared to (Gödde and Kurz, 2001) we get a
good match. The data is from a single simulation run. Data from the
initial vasculature and from the tumor vasculature at \timeis{1000}
are both displayed in the same plots.

For the initial network we generally observe that the variance 
of the flow related parameters increases drastically towards the 
capillaries. While this might be an artifact of the artificially 
constructed network, averaged values show physiologically 
sound characteristics.

The dilation of tumor internal vessels leads to samples clustered at
the maximum dilation radius ($\paramRMax=25$\micron) and therefore one
observes larger ranges of hemodynamic variables. In particular, the
blood pressure in tumor vessels ranges from typical arterial values to
venous values as these vessels connect both sides via shunts by a
continuously varying pressure potential. Consequently the pressure in
arteries, over which the tumor as grown, is lower than normal while
the pressure in former veins is elevated.

In the flow computations we fix the pressure difference between
supply-and drain, which means that the flow rates in the major
vessels are determined by the total resistance of the network.
Lowering this resistance by arterio-venous shunts increases the 
blood throughput, which is evident in \figinline{10}c when 
observing the increased flow rates in the thickest vessels.
This increase is in particular pronounced on the venous side. 
The flow rate in the initial network grow slightly stronger with the radius
on the arterial side than on the venous. Therefore the difference
between tumor network and normal network is less pronounced for
arteries. Interestingly many of the dilated capillaries
at $\paramRMax$ fall below the normal average flow rate.

In contrast, most non-dilated small capillaries which are obviously
located in the outer rim fall in the same ranges than normal
vessels. A few vessels though show significantly lowered flow and
velocity values. In \sref{SEC:DRUG} we study the impact of them upon
drug delivery.

\subsection{Parameter dependencies}
\label{SEC:PARAMETERVARI}

Here we discuss the role of the most sensitive parameter dependencies
and model changes. The large shear force variations among capillaries
in the original vasculature leads us to the question whether an
absolute threshold for the collapse criterion is
appropriate. Therefore we also checked the result of setting
$\paramProbColl \propto 1.0-(f/f_{init})/\paramFCrit$ for
$f/f_{init}<\paramFCrit$, where $f_{init}$ is the shear force in the
of the original vasculature. The result is less MVD fluctuations in
the samples and also locally a more homogeneous distribution.

In the present model variant, vessels can collapse any time after
degradation ($\idWall<\paramWallCrit$), depending on shear-stress. It
is however also an option to restrict vessel collapses close to the
tumor boundary. For example, it has been suggested that the death of
tumor cells releases solid pressure from nearby vessels which is
otherwise exerted by the tumor (Griffon-Etienne et al., 1999). In
previous papers we assumed the existence of a stable radius,
preventing collapses in the center where vessel radi are
larger. Indeed, the EphB4 signaling mechanism which is related to
circumferential growth, can also lead to reduced leakiness, tightened EC
junctions and increased endothelium/pericyte interaction (Erber et~al., 2006),
making an actual stability improvement plausible.  Again, we have
implemented this very simplistically by modulating the collapse
probability: $\paramProbColl = \paramProbCollRing = \paramProbCollFull
\cdot\Theta(x-\tumradius-\paramFCritDelta)$, where $\Theta$ is the 
Heaviside function, $\tumradius$ the radial tumor extend, and
$\paramFCritDelta$ is the ring width measured from the invasive
edge. The result obviously depends strongly on the time vessels spend
in the collapse region, given implicitly by $\paramFCritDelta$,
$\paramWallDecreaseRate$, the radius $r$ and the tumor expansion
rate. If this time is of the order of the mean survival time
$\deltat/\paramProbColl$ a transition occurs toward a highly
vascularized center, reminiscent of the ``percolation transition''
analyzed previously (Bartha and Rieger, 2006; Welter et al. 2007; Lee
et al., 2007). In this case the random collapse-process also plays a
more dominant role because many weakly-perfused vessels can survive
whereas, using $\paramProbCollFull$, these vessels would eventually
regress if one just waits sufficiently long.
\figinline{11} shows respective simulation results, 
using $\paramFCritDelta=400$\micron, $\paramFCritN=3$Pa and
all other parameters unchanged.
One can see increased formation of high-MVD hot spots and less
isolated threads than for example in \figinline{5}.
The MVD is slightly higher, but easily tunable by $\paramProbCollN$
and $\paramFCritN$.
Generally, $\paramProbCollRing$
also leads to less fluctuations with respect to initial networks
apparently because initially thick vessels provide a backbone which
never collapses due to $\idWall>\paramWallCrit$ while being in the
``collapse ring''. 

By default sprouting is allowed from veins and capillaries but not
from arteries.  Enabling sprouting from all vessels naturally leads to
ca. 40 \% higher MVD in the boundary, while the central MVD also
depends on the collapse model.  Using parameters with
$\paramProbCollRing$ so that the effect of $\paramProbCollRing$
becomes significant, i.e. near a transition to fully vascularized
tumor, the MVD will increase, but otherwise not.
Because the boundary becomes more homogeneously vascularized also the
\Otwo field becomes more homogeneous. But with the current \Otwo proliferation 
threshold $\tcProlOxy \approx 0.9\meanval{\foxy}$ this does not alter
the tumors expansion rate due to TCs proliferating around smaller
low-\Otwo regions and enclosing them. We expect this to be different
in reality. TCs would likely migrate into low-\Otwo regions driven by
the pressure of the surrounding tissue where they remain in a
quiescent state.

\subsection{Drug flow}  
\label{SEC:DRUG}

In order to assess the effects of typical tumor network morphology on
transport of drugs into and through the tumor, we analyze the
time-dependent concentration distribution of drug during a continuous
injection into the blood stream, starting at arterial boundary nodes.

The computational model for drug flow in a vascular network is
described in detail in (Welter et~al., 2007; McDougall et~al.,
2002). The starting point is a given configuration for the vasculature
in our model with precomputed/prescribed variables for flow, flow
velocity, segment length, and radius of the pipes in the network. In
addition, a mass parameter $m$ is now associated with segments
describing the amount of drug in the segments volume.  The mass
content $m$ is deterministically updated in successive time steps as
follows: First the drug amount flowing out of segments is determined
and added to respective node-mass variables.  Under the assumption of
perfect mixing, the nodal masses are then redistributed into further
downstream segments. Thereby mass conservation is strictly enforced. A
detailed description can be found in (Welter et~al., 2007). The most
severe limitation of this model is that there is no exchange with
extra vascular space and therefore also no uptake by the tumor.

The results in the following were obtained with an continuous
injection into the vasculatures from the case $\paramFCritN$ = 1. From
t=0 on, blood with solute drug at conc. $\paramDrugInit$ = 1 flows
into the vasculature which is initially filled with ``clean'' blood. We
don't consider bolus injections because the flow rates are of the
order of mm/s, which is sufficient to saturate 80-100\% of the vessels
within seconds, depending on the network configuration.

\figinline{15} shows a sequence of snapshots from configuration
(a) \timet=0 .. 6.6s. Drug enters the system via the arteries and
flows downstream with a sharp transition at the drug/clean
interface. When vessels merge in upstream direction mixing with clean
blood occurs and so the concentration increases moderately in the
veins.  After 20 s near full saturation is achieved. Dilation of tumor
internal vessels and apparent direct connections to feeding arterioles
lead to comparably fast filling with drug, whereas a few regions in
the highly vascularized boundary take significantly longer to be
filled.  Depending on the exact network configuration we have
consistently encountered such strong variations in drug delivery to
the outer rim. It is likely to be caused by the shunts which lead to
decreased flow rates in the surrounding tissue.
\figinline{10}c also shows the existence of capillary sized
tumor vessels with very low flow rates of the order of $10^{3}$\micron$^3$/s,
corresponding to an average flow velocity of the order of 10\micron/s.
The average flow velocity over all vessels in the original network is
$\sim 1.8$mm/s.

Quantitative analysis of drug efficiency impact is only possible to a
limited degree due to the lack of tumor uptake modeling and because
even few surviving TCs can grow a new tumor mass. None the less we
measured statistics for the total amount of drug in the
vasculature. \figinline{16}a for example, shows the fractional
length of the tumor vessel-network for which the maximum
drug-concentration was larger than indicated on the
``$c$''-axis. After 20 s, on average 94 $\pm 3\%$ of the vasculature
had been exposed to a drug concentration of at least 80\%.  For higher
concentrations the curve decreases drastically. It also shows
evidently by the drop at c=0 that $4 \pm 3\%$ had not been drug
perfused at all. In absolute numbers the latter corresponds to $15 \pm
12$mm. \figinline{16}b shows the fractional length of the tumor
vessel -network for which the concentration was larger than the
threshold $c=0.25$ for a total duration greater than indicated on the
x-axis as $t_e$. This measurement was also done at t=20s, thus the
exposure time cannot be longer than that. One can see, for example,
that $93\pm 4\%$ of the vasculature had been exposed 50\% drug
concentration longer than 10s. Considering the flow rates/velocities
mentioned above one can expect the remaining few millimeters to be
drug perfused within a time frame of minutes. We therefore still
believe that the geometry and hydrodynamic flow characteristics do not
pose an inherent problem to drug delivery.

\section{Hot-spots and spatial inhomogeneities}

Already by visual inspection one observes a coincidence of the
location of hot spots in the tumor vasculature with the location of the
thicker vessels. For example in configuration (a) where an artery is
close to a vein, or vice versa, an increased number of vessels survive
in between. On the other configuration similar behavior can be observed
on smaller scale. The blood pressure in thick vessels is approximately
constant compared to the steep drops in the capillary bed. This is due
to the orders of magnitudes larger flow conductivity. It is plausible
that in (a) for example, a high global vertical pressure drop leads
to high shear forces in between the two major vessels and therefore
increased survivability. Furthermore the asymmetry in the shear stress
distribution over the vascular trees (meaning that given vessels of a
certain radius, the shear stress in arteries is usually higher than in
veins; see \figinline{10}), apparently leads to hot spot formation
preferably close to the arterial branches.

This motivates a quantitative analysis of the relation between
inter-vascular ``blood-pressure'' gradients in the original
vasculature and the local MVD of the tumor vasculature. This is done
as follows: Since new vessels formed during the process of sprouting
angiogenesis will experience a shear force proportional to the pressure
drop from one end to the other. Therefore we first compute a pressure
field $P(\vecr)$ as solution to the Laplace equation $\Delta P = 0$
subject to the boundary condition that $P$ is set to the blood
pressure at vessel sites. The resulting field interpolates pressures
between adjacent vessels, and would change perfectly linear between
two infinite parallel vessels.  At any given point, we can obtain the
magnitude of the gradient , which we use to compute a spatial average
of $\left\|\nabla P\right\|$.  We do this for \timeis{0}. The
resulting mean value $\meanval{\left\|\nabla P \right\|}$, is plotted
together with an estimate for the tumor MVD at \timeis{1200} (defined
as the fraction of occupied sites) as displayed in \figinline{12}. The
main plot shows the MVD versus $\meanval{\left\|\nabla P \right\|}$
over entire tumors with one data point per simulation run. The
correlation coefficient is usually very large $\approx 0.9$ for this
global measurement.  Analogously we analyzed single tumor instances
where the averaged is taken locally over randomly distributed small
regions (discs with 150\micron radius). The inlet shows the resulting
distribution of one such measurement. The correlation coefficient in
this instance is already low ($\approx$ 0.4). On average we observed
it to vary with the model details and even with the initial
vasculature between 0.2 and 0.4.

\figinline{13} is a visualization  of the statistical correlation
observed in the above analysis. We take the data fro the system
configuration in the simulation run shown in \figinline{5}. The bottom
image (c) shows the site occupation by the vessel network. It is easy
to identify the tumor network by its typical structure. In (a), where
$\left\|\nabla P \right\|$ is displayed, one observes that indeed the
local gradients are stronger in the area close to the major artery in
the top half, than close to the vein in the lower half.  The snapshots
on the right column (b) and (d) show respective mean value samples of
the left hand side. These distributions show the behavior that zones
with elevated $\meanval{\left\|\nabla P \right\|}$ are also probable
locations of high MVD.

A plausible argument for the existence of the correlation between
hydrodynamic characteristics of the initial network and local
morphological features in the evolved tumor vasculature is the
following: In regions with originally strong inter-vascular pressure
drops, tumor vessels would be less prone to collapse because sprouting
forms new connections whose shear-stress is proportional to $\|\nabla
P\|$. Thus, shear-stress stabilization would prevent regression in
high-$\|\nabla P\|$ regions.  Since single collapse events lead to
long-ranged collapses of adjacent network sections, we believe that
therefore local measurements hardly show correlations.  Note that the
intra-vascular blood-pressure at in-and outflow vessels is prescribed
and depends on the vessel radius. Therefore only the positioning of
initial vessels leads to the variety of emerging tumor morphologies.

We further quantify the spatial inhomogeneities by probability
distributions for local MVD (\figinline{14}a), necrotic
region size (b) and hot spot size (c). The distributions are estimated
via histograms of the respective local measures, whereby we insert
data from multiple (40) runs into the same histogram in order to get a
smooth curve with reasonably small bin. The MVD in
\figinline{14}a is estimated as fraction of occupied sites
within boxes (250\micron in size) of a regular grid. The plot displays
two peaks. One large at low MVD$\approx$0.07 which decays
algebraically with exponent $\sim 1.4$ till smaller peak at
MVD$\approx$0.45. The latter stems from the high-MVD zone in the tumor
periphery. Not included is a peak which appears at MVD=0 due to large
non-vascularized regions. The necrotic-region size distribution in
\figinline{14}b plots the probability to find a connected
cluster of dead TCs with a certain volume, i.e.  number of cells.  The
hot-spot size distribution in \figinline{14}c plots the
probability to find a connected cluster of a certain number of sites
where the local MVD is larger than a threshold value (here
0.15). Therefor we compute an estimate for the MVD for each lattice
site as the fraction of occupied sites within a 250\micron diameter
disc. Both of the latter distributions exhibit purely algebraic decay
also with the exponent $\sim 1.4$.

The fractal dimension $d_f$ is often used to quantify tumor networks
and has been found capable to distinguish them from normal vascular
networks (Baish and Jain, 2000).  Therefore we do box counting
analysis, which is usual practice to estimate $d_f$ for natural
objects. Perfect fractals show a power law $N \propto
\epsilon^{-d_f}$ in the number of touched boxes $N$ vs. size $\epsilon$, 
yielding a constant slope in a log-log plot. We typically observe a
parabolic shape of the local slope curve with a small constant plateau
over half a decade. Due to the problems associated with $d_f$
estimates, such plateaus are often considered sufficient to extract
$d_f$; see the discussion in (Chung and Chung, 2001). In this context
one also speaks of box counting dimension.  On average we obtain
$d_f=1.76\pm0.03$ for the final internal tumor network; $d_f=1.79
\pm0.03$ including high-MVD peripheral region. The error is given by
the root mean square deviations among 40 simulation runs with the same
parameter set. Fluctuations of this magnitude are very significant,
which is why we consider these results as unreliable and also show no
data here. Furthermore resembling previous results, we still find our
$d_f$ estimate being primarily determined by the MVD.

\section{Discussion}
\label{SEC:FINAL}

The conclusions that can be drawn from the theoretical model for
vascular remodeling of an arterio-venous network during tumor growth
are manifold: As has already been conjectured on the basis of an
analogous model with a regular or lattice-like blood vessel network
(Welter et al., 2006) the global characteristics of the emerging tumor
vasculature is expected to be independent of the specific details of
the initial vasculature: the resulting morphology is compartmentalized
into a highly vascularized tumor perimeter, a tumor periphery with
large vessels density and dilated vessels and a central region
containing necrotic regions with a low micro-vascular density threaded
by extremely dilated vessels. The basic mechanisms leading to this
morphology are identified as sprouting angiogenesis in the tumor
perimeter, a switch of the vascularization program to circumferential
growth plus vessel regression within the tumor.

On top of these global feature our model predicts spatial
inhomogeneities, visible in drastic MVD variations. These are strongly
correlated with the local characteristics of the initial
arterio-venous vessel network, which are absent in a regular
homogeneous initial vasculature. In simulation runs with identical
parameters but different arterio-venous network configuration, cases
can be observed ranging from tumors threaded mostly by individual
vessels, to tumors that exhibit many dense clusters connected by few
short chains. The reason for this variability the asymmetry of the
shear force distribution between the arterial and venous side of 
the vascular trees.

Depending on the details of the initial network in our model the tumor
vasculature develops isolated highly vascularized clusters connected
by thick vessels. These ``hot spots'' are also observed in real tumor
and serve as an important diagnostic tool in cancer therapy. Already
by a visual comparison of the starting network with the final tumor
network one observes that hot spots form more frequently in those
regions where the starting network contained predominantly arteries.
We therefore analyzed the correlation between various local
hydrodynamic quantities of the original network (blood pressure, blood
pressure gradient, blood flow) and the local MVD in the tumor network
and found a significant correlation between local blood pressure
gradient of the original arterio-venous network and the most probable
locations of hot spots in the tumor vasculature. This is plausible
since a high pressure gradient within the vessels implies a high shear
force exerted by the blood on the vessel walls, which stabilizes the
vessel against collapse and therefore leads to an increased vessel
survival probability when the tumor has grown over this region. Since
we expect these mechanisms also to be at work in real tumors it should
therefore in principle be possible to predict on the basis of an
analysis of the blood vessel network in the healthy tissue, where most
likely the tumor vasculature develops its hot spots. We note that in
our model this hydrodynamic mechanism plays a dominant role in the the
hot spot formation, whereas potential local variations in pro- and
anti-angiogenic effectors within the tumor were not involved but
could also play a role in this process.

Remarkably an association between the MVD of the original vasculature
in normal tissue and hot spot vascular density in the tumor has been
reported for carcinoma (H\"ockel et al., 2001). Here the MVD of normal
tissue was measured far away from the tumor and it was found that it
correlates well with the MVD of the hot spots in the tumor. Our
interpretation of this result is that sprouting angiogenesis is not
strong in these tumor samples (in fact the MVD in the tumor periphery
was increased by only ca. 20\%) and hot spots in the tumor represent
regions of the original vasculature that could survive vessel collapse
due to locally increased pressure gradients. For a reliable check of
our hypothesis one would have to analyze the pressure gradients in the
vasculature of the normal tissue {\it before} the tumor grows over it,
i.e.\ under experimental conditions of an implanted tumor for instance.

Various probability distributions that quantify the spatial
inhomogeneities of the tumor network in our model turn out to display
an interesting behavior, too: The probability distribution of local
MVD values as well as the probability distribution of the hot spot
volume (defined as the size of connected clusters of regions with MVD
larger than the original MVD) have an algebraic tail with exponent
$\sim 1.4$. Also the size distribution of the connected clusters of
necrotic tumor regions showed an algebraic tail (again with exponent
$\sim 1.4$). These power laws indicate a critical state in accordance
with the fractal properties of the tumor network: It implies the
absence of a particular lengths scales over which size distribution
would decay exponentially and resemble the size distribution that one
encounters in percolation at the critical point, the percolation
threshold.

When analyzing the blood flow characteristics in emerging tumor
vasculature of our model we found an interesting phenomenon which is
naturally absent in models with a regular starting configuration:
Thick arterioles and venoules provide a well conducting support
structure around the tumor. Since the total pressure difference
between the tree roots is fixed, the transported blood volume is given
by the total flow resistance of entire vascular tree. Dilation of a
few selected vessels forming a path between the tree roots can remove
a bottlenecks formed by thinner vessels. Angiogenesis which provides
additional vessels thereby facilitates the creation arterio-venous
short-cuts, or shunts, through multiple partly disjoint paths. This
leads to an increased blood flow through the tumor vasculature when
compared with the starting vasculature. In contrast to this in a
regular network the total flow resistance is dominated by the network
outside of the tumor (Bartha and Rieger, 2006, Welter et al. 2007) and
the flow cannot not increase via the dilation of tumor internal
vessels. There are experimental studies which agree with the
predictions of our model. In (Sahani et al., 2005) perfusion parameters
in rectal cancer were measured via computer tomography where
consistently increased blood flow is reported by approximately a factor
of two compared to normal tissue. There it is also argue that
angiogenesis facilitates the creation of arterio-venous shunts which
bypass the capillary network - the same mechanism by which blood
increases in our model.

These arterio-venous shunts should also lead to an increased blood
pressure range in tumor vasculatures: They connect vessels with high
pressure to vessels with low pressure, which then varies continuously
within the connecting vessel. Indeed we find a wide spread among
pressures over a short range of radii near $r_{max}$ in our model,
which is somewhat in contrast with the prevailing view that the
pressure as function of vessel radius is reduced in the venous part in
tumors (Jain, 1988). Our result indicate that the measure
used to assess the relation of tumor blood flow to normal tissue has
critical impact on the results. It has been shown that measuring
perfusion as function the spatial location can give rise to elevated
flow rates, whereas comparison of flow rates between vessels of a
certain radius may show either increased as well as decreased flow
rates depending on the radius and the location in the vascular
hierarchy.

Our drug flow computations again showed that a drug bolus, which is
injected into the source arteries of the blood vessel network in our
model for some time, reaches all perfused blood vessels, although the
thin capillaries in the highly vascularized tumor boundary needs a
longer time to become filled with drug. Blood-borne delivery of
therapeutics into the tumor-vascualture does not appear to be an
obstacle for a successful chemotherapy. The reasons for failure of
drug delivery to tumor cells are more likely related to drug uptake
the drug transport through the tumor tissue (Minchinton and Tannock,
2006). Since blood is in contact with the interstitial fluid, due to
leaks in tumor vessels, fluid exchange can take place with the
ECM. This has been shown to lead to steady states with elevated IFP
levels, where the difference between micro-vascular pressure (MPD) and
IFP generally lower than normal (Boucher et al., 1996). 
Since convective transport is driven by pressure
differences, high IFP could pose a barrier to drug delivery (Hassid
et~al., 2006)). On the other hand, leakiness and MVP-IFP
gradients could lead to premature release predominantly in locally
restricted regions around vessels where blood enters the
tumor. Vessels in the outflow regions would thus be depleted of
drug. Locally released drug would then be transported by IFP gradients
out of the tumor. Furthermore drugs usually consist of large
macromolecules. Their low diffusibility through the vessel wall and
generally lower diffusibility than \Otwo could lead to situations
where sufficient \Otwo reaches certain TCs to let them remain viable,
but not enough drug reaches them to kill them off due to the lower
diffusion radius.

In future work it would be useful to study the remodeling of a three
dimensional arterio-venous network. Getting access to a realistic
initial network is likely to pose a hard problem. In particular if it
must be artificially created. Otherwise the extension to 3d is
straight forward and has already been done in (Lee et al., 2006)
with similar results. Further aspects to assess are interstitial
fluid transport and $O_2$ release with associated intra-vascular $O_2$
decrease. Furthermore implementation of a tumor model which supports
quantitative assessment of stresses in the tissue appears mandatory
for studies of irregular tumor morphologies.

\vfill
\eject

\newpage

\section*{References}

Bartha, K., Rieger, H., 2006. Vascular network remodeling via 
vessel cooption, regression and   growth in tumors. J. Theor. 
Biol. 241, 903-918.
\\ 

Baish, J.~W., Jain, R.~K., 2000. Fractals and Cancer. Persp. 
Cancer Res. 60, 3683--3688.
\\

Belien, J.A., Somi, S., de Jong, J.S., van Diest, P.J., Baak, J.P.,
1999. Fully automated micro-vessel counting and hot spot selection
by image processing of whole tumor sections in invasive breast cancer.
J. Clin. Path. 52, 184-192.
\\

Carmeliet, P., Jain, R. K., 2000. Angiogenesis in cancer and other
diseases. Nature, 407, 249-257.
\\

Carmeliet, P., Tessier-Lavigne, M., 2005. Common mechanisms of nerve
and blood vessel wiring. Nature 436, 193-200.
\\

Cassot, F., F.~Lauwers, C.~F., Prohaska, S., Lauwers-Cances, V., 2006. 
A Novel Three-Dimensional Computer-Assisted Method for a Quantitative 
Study of Microvascular Networks of the Human Cerebral Cortex. 
Microcirculation 13, 1-18.
\\ 

Chung, H.-W., Chung, H.-J., 2001. Correspondence re: J. W. Baish and
R. K. Jain, Fractals and Cancer. Cancer Res., 60: 3683-3688. Cancer
Res. 61, 8347--8351.
\\

Döme, B., Paku, S., Somlai, B., Tímár, J., 2002. Vascularization 
of cutaneous melanoma involves vessel co-option and has clinical 
significance. J. Path. 197, 355-362.
\\ 

Erber, R., Eichelsbacher, U., Powajbo, V., Korn, T., Djonov, V.,
Lin, J., Hammes, H.-P., Grobholz, R., Ullrich, A. Vajkoczy, P., 
2006. EphB4 controls blood vascular morphogenesis during postnatal 
angiogenesis. EMBO 25, 628-641.
\\ 

Folkman, J., Bach, M., Rowe, J.W., Davidoff, F., Lambert, P., Hirsch,
C., Goldberg, A., Hiatt, H.H., Glass, J., Henshaw, E., 1971. 
Tumor angiogenesis therapeutic implications.
N. Engl. J. Med. 285, 1182-1186.
\\

Folkman J., 1990. What is the evidence that tumors are angiogenesis 
dependent. J. Natl. Cancer Instit. 82, 4-6.
\\

Gödde, R., Kurz, H., 2001. Structural and Biophysical Simulation 
of Angiogenesis and Vascular Remodeling. Dev. Dyn. 220, 387-401.
\\ 

Hassid, Y., Furman-Haran, E., Margalit, R., Eilam, R., Degani, H.,
2006. Noninvasive Magnetic Resonance Imaging of Transport and
Interstitial Fluid Pressure in Ectopic Human Lung Tumors. Cancer
Res. 66, 4159-4166.
\\ 

Hannahan, D., Folkman J., 1996. Patterns and emerging mechanisms 
of the angiogenic switch during tumorigenesis. Cell 86, 353-364.
\\

H\"ockel, S., Schlenger, K., Vaupel, P., H\"ockel, M., 2001.
Association between host tissue vascularity and the prognostically 
relevant tumor vascularity in human cervical cancer.
In. J. Oncol. 19, 827-832.
\\

Holash, J., Maisonpierre, P.~C., Compton, D., Boland, P., Alexander, 
C.~R., Zagzag, D., Yancopoulos, G.~D., Wiegand, S.~J., 1999a. Vessel 
Cooption, Regression, and Growth in Tumors Mediated by Angiopoietins 
and VEGF. Science 284, 1994-1998.
\\ 

Holash, J., Wiegand, S., Yancopoulos, G., 1999b. New
model of tumor angiogenesis: dynamic balance between vessel
regression and growth mediated by angiopoietins and VEGF. Oncogene
18, 5356-5362.
\\

Kullander, K., Klein, R., 2002. Mechanisms and functions of Eph and
ephrin signalling. Nature Rev. Mol. Cell. Biol. 3, 475-486. 
\\

Mandelbrot, B.B., 1982. The fractal geometry of nature.
Freeman, New York.
\\

McDonald, D.~M., Choyke, P.~L., 2003. Imaging of angiogenesis: 
from microscope to clinic. Nature Med. 9, 713-725.
\\ 

Minchinton, A. I., Tannock, I. F., 2006. Drug penetration in solid tumours.
Nature Reviews Cancer 6, 583-592.
\\

Mironov, V., Little, C., Sage, H. (Eds.), 1998. Vascular Morphogenesis: 
in Vivo, in Vitro, in Mente. Birkhäuser Boston.
\\ 

Murray, C.~D., 1926. The physiological principle of minimum work: The 
vascular system and the cost of blood volume. In: Proc. Natl. Acad. 
Sci. USA, vol.~12, 207-214.
\\ 

Pahernik, S., Griebel, J., Botzlar, A., Gneiting, T., Brandl., M.,
Dellian, M., Goetz, A. E., 2001. Quantitative imaging of tumour blood 
flow by contrast-enhanced magnetic resonance imaging. Brit. J. Canc. 
85, 1655-1663.
\\

Paku, S., 1998. Current concepts of tumor-induced
angiogenesis. Pathol. Oncol. Res. 4, 62-75.
\\

Pries, A.~R., Reglin, B., Secomb, T.~W., 2005. Remodeling of Blood Vessels: 
Responses of Diamter and Wall Thickness to Hemodynamic and Metabolic Stimuli. 
Hypertension 46, 725-731.
\\ 

Pries, A.~R., Secomb, T.~W., Gaehtgens, P., 1995. Design principles of 
vascular beds. Circ. Res. 77, 1017-1022.
\\ 

Schreiner, W., Buxbaum, P.F., 1993.
Computer-optimization of vascular trees.
IEEE Trans. Biomed. Eng. 40, 482-491.
\\

Welter, M., Rieger, H., Bartha, K., 2007. Emergent vascular network 
inhomogeneities and resulting blood flow   patterns in a growing tumor. 
To be published in J. Theor. Biol.
\\ 

Weidner, N., 1995. Current pathological methods for measuring
intratumoral microvessel density within breast carcinoma and other
solid tumors. Breast Canc. Res. Treat. 36, 169-180.
\\

\newpage
\section*{Tables}

\textbf{Table 1:}
Listing of model parameters with values for our base case simulation
runs.
\bigskip
\begin{center}
\renewcommand{\arraystretch}{1.5}
\begin{tabular}{|>{$}r<{$}|r|r|r|}
\hline
\texthack{Parameter} 								& Value 									& Description 									& Reference	 											\\
\hline
\hline
\latticeconstgen     & 60 \micron     &   Lattice const. (Tree-generator) &  \\
\hline
\latticeconst							& 10 \micron 							& Lattice const. (Tumor-model)								&  																\\
\hline
\latticesites 						& 1200                    & Lattice size (Tumor-model)       &                         				\\
\hline
\deltat									& 1 h                     & Time step                     &																	\\
\hline
\hline
\paramTumInit             & 1000                    & Initial number of TCs                  &          \\
\hline
\hline\oxyqN										& 0.02 									  & \Otwo source coefficient      & (Secombet~al. 2004) 						\\
\hline
\foxyktiss										& (80 \micron)$^{-2}$ 		& consumption in normal tissue  & (Carmeliet and Jain, 2000)  		\\
\hline
\foxyktum											& $4 \foxyktiss$ 							& consumption tumor tissue			&  																\\
\hline
\tcProlOxy 							& $0.3 \approx 0.9\meanval{\foxy}$& TC \Otwo prol. threshold    &          \\
\hline
\foxyb								 & 0.45 & dimensionless blood-\Otwo level & \\
\hline
\tcDeathOxy              & $\tcProlOxy/10$         & TC hypoxia threshold          &          \\
\hline
\hline
\paramRGf							& 200 \micron & Growthfactor diffusion radius &  (Nehls et~al. 1998)  \\ 
\hline
\vessProlGf							& $10^{-4} $              & sprouting GF threshold        &          \\
\hline
\hline
\paramTimeUO            	& 100 h                   & TC survival time under hypoxia&   (Yu et~al. 2002)  \\
\hline
\paramTimeTcProl 				& 10 h         & TC proliferation time &\\
\hline
\paramTimeSprout 			& 5 h         & Sprout initiation/extension time &\\
\hline
\paramTimeDil 				& 40 h 				& Vessel dilation time &  (Döme et~al. 2002)\\
\hline
\paramTimesprmax 				& 100 h  				& Time till sprout regression & (Nehls et~al. 1998)\\
\hline
\paramSproutParentInTumor & 25 h & Time till switch to circumferential growth & \\
\hline
\paramWallDecreaseRate & 0.04 \micron/h & Wall thickness/stability decrease rate &  \\  
\hline
\hline
\paramRInit						& 4 \micron  & Initial vessel radius &  \\
\hline
\paramRMax             & 25 \micron    & Max. vessel radius & (Döme et~al. 2002)\\
\hline
\paramLenSprout 				& 20 \micron 	& Min. distance between junctions &(Döme et~al. 2002) \\
\hline
\paramFCritN 					& 1 Pa			& Peak critical shear force &	\\
\hline
\paramProbCollN				& 0.1         & Peak collapse probability &      \\
\hline
\end{tabular}
\end{center}
\vfil


\newpage
\section*{Figure captions}

\textbf{\figinline{1}:}
Illustration of the model state: All discrete elements are aligned
at the triangular lattice shown in the background. Blue bars represent
vessels occupying bonds. The set containing all vessels is denoted by
$\vesselset$. One vessel is highlighted by a black frame,
it's junctions to other vessels are denoted by $\vecr_1$.$\vecr_2$.
Tumor cells (TCs) can occupy lattice sites with a one to one
relationship. The set containing TCs is denoted as $\tumset$.
One TC is displayed as yellow hexagon.
\bigskip

\textbf{\figinline{2}:}
Sketch of the dynamical processes in the model: 
(a) TC Proliferation, (b) TC Death, (c) Sprout
formation, (d) Sprout migration, (e) Vessel
regression and removal, (f) Dilation and wall
degradation. See text for details.
\bigskip

\textbf{\figinline{3}:}
Examples of starting configurations of the model at \timeis{0}. 
We chose four different basic layouts for the vascular
network denoted as A,B,C,D. They differ in the placement and lengths
of major supply-and drain vessels, which are randomly distributed.
This is indicated in the top right inlets. Arteries are red,
veins blue, black bars show intervals from which starting
locations are drawn while the wedge shaped part of the red/blue
bars indicate intervals from which starting lengths are drawn.
The color code of the resulting networks shows the blood pressure
distribution from 2kPa (red) to 12kPa (blue). The small yellow
area in the system center is the tumor nucleus initialized by
eden growth to maximum of 1000 TCs. The background in normal
tissue/extracellular matrix is colored green if the
growthfactor is sufficient to enable sprouting. Otherwise those
areas are white.
\bigskip

\textbf{\figinline{4}:}
Snapshots of the dynamical evolution at times 
\timeis{0, 200, 600} of the model in a single simulation run
starting in configuration A (shown in Fig. 3a) at $t=0$. Viable TCs
are bright yellow, while necrotic regions are dark yellow.  Vessels
are color coded by blood pressure as in \figinline{3}.  The background
in normal tissue/extracellular matrix is colored green if the
growthfactor is sufficient to enable sprouting. Otherwise those areas
are white.
\bigskip

\textbf{\figinline{5}:}
Final configuration at \timeis{1200} for the simulation run depicted
in Fig.4. The color code is identical to \figinline{4}. Note how
vessels in the tumor center serve as arterioveneous shunts. They have
dilated diameters and thus carry high blood throughput.  The formation
of hot spots with increased local MVD is also evident.
\bigskip

\textbf{\figinline{6}:}
Final configurations at (\timeis{1200}) of simulation runs of the
model startig with the initial configurations B, C, D shown in
\figinline{3}. The color code is identical to \figinline{4}. The
formation of hot spots is clearly visible in the central region in B,
but not so in C and D. There increased MVD can be observed in a few
locations close to the tumor boundary.
\bigskip

\textbf{\figinline{7}:}
Magnified view of the tumor boundary showing in detail
how the vasculature is remodeled close to the invasive 
edge. Alternating with very high MVD on small scale, there
are ``holes'' of the size of the intervascular distance
in normal tissue. This is an effect of prohibiting sprouts
from arteries, which would otherwise fill the holes, leading
to a homogeneous vascular density. Behind the peripheral
zone, the MVD drops rapidly. One can further see a few
isolated dilated vessels directly connected to 
arterioles/venoules in healthy tissue beyond the invasive edge.
For the color code see the caption of 
\figinline{4}.
\bigskip

\textbf{\figinline{8}:}
Flow rate distribution in the tumor vessel network depicted in
\figinline{5}. The color code is on a logarithmic scale as indicated
on the left side. The flow rate here is the blood volume per time
through the vessel cross-section, closely related to the blood flow
per tissue volume as commonly measured in experiments. Dilated
tumor-internal vessels usually exhibit high flow rates.  If one would
measure in/outflow per tissue volume element as in dynamic MRT
(magnetic resonance tomography) measurements (Pahernik et al., 2001)
one would observe hot-spots located at the respective high-MVD zones.
\bigskip

\textbf{\figinline{9}:}
Radial distributions of various quantities characterizing the
dynamical development of the compartmentalization of the tumor
vasculature in the model: The radius $r$ here denotes the distance
from the tumor center. One data point represents an average over a
concentric 50\micron wide shell, and over 40 simulation runs, 10 per
configuration A,B,C,D (see text or \figinline{3}).  (a) shows the
micro vascular density, (b) the \Otwo level and (c) the tumor cell
density. The right column shows blood flow related variables: (d) the
vessel radius, (e) the flow rate and (f) the shear force, which are in
contrast to (a),(b),(c) given as the average over actually occupied
sites within a shell, instead of over all sites. For \timeis{1200}
also the local variations (root mean square deviations) are indicated
as error bars.
\bigskip

\textbf{\figinline{10}:}
Scatter plots showing hydrodynamic quantities for individual vessel
segments as a function of the vessel radius: (a) blood pressure, (b)
wall shear force, and (c) flow rate through the vessel cross
section. Each data point is a sample from a randomly chosen position,
uniformly distributed over the vascular network. Samples from both,
the initial vasculature at \timeis{0}, and the tumor tumor network at
\timeis{1000} are shown. For the former, arteries are displayed in red
with negative radius, veins are blue and capillaries are pink. Tumor
vessels that were initially arteries are displayed yellow, those that
were veins are light blue. New vessels and capillaries are black.
Dilation causes the latter vessels to span radius values from $5$ to
$25$\micron (the maximum dilation radius $\paramRMax$), but due to the
lack of hierarchical organization in the tumor they cannot be
contribute to arteries or veins.  Therefore respective data points
were put with probability 1/2 on one or the other side.  Note that
tumor vessels capped at $\paramRMax$ display a much larger range of
values than normal (see text). Also note that (b) and (c) are
presented as linear-log plots.
\bigskip

\textbf{\figinline{11}:}
Final configuration at \timeis{1200} of a simulation run 
using an alternate vessel collapse rule where vessel 
collapses are restricted to a thin band (here 400\micron
in width) behind the invasive edge. The collapse probability 
$\paramProbCollRing$ is thus
modulated with an appropriate term which depends on the distance
to the tumor center (see text). In this instance, parameters
are such that the mean survival time of instable
vessels is of the order of the time they spend in the 
``collapse region'', leading to a dominant role of the
random collapse process for the resulting morphologies,
and putting the system near a transition to a fully
vascularized tumor. The parameter set is the same as
in the base case, except for the critical shear force
$\paramFCritN$ which is set to 3Pa, instead of 1Pa.
\bigskip

\textbf{\figinline{12}:}
Local MVD in the tumor vasculature versus local pressure gradient
$\meanval{|\Delta P|}$ (i.e.\ blood pressure differences between
neighboring vessels) in the starting vasculature. In the main plot
data points are averaged over different runs of the base case scanrio,
the averages are taken over the entire tumor interiors
(excluding the vascularized boundary region) at
\timeis{1200}. The inlet is a scatter plot for a single simualtion run, where
the an average is performed locally over a small disc with radius
150\micron (see \figinline{13}).
\bigskip

\textbf{\figinline{13}:}
Digitized data processing of network configurations demonstrating the
correlation between the intervascular pressure gradient in the
starting network with the local MVD in the tumor network.
(a) Local magnitude of the ``intervascular pressure'' gradient
$\|\Delta P\|$ in the initial network (see text). (b) Smoothing of the
data in a: locally averaged values of $\|\Delta P\|$ were computed at
randomly distributed locations, each by averaging over a
150\micron-radius disc, then the space closest to some data point is
filled with a gray scale value proportional to $\meanval{\|\Delta
P\|}$.  (c) Local MVD of the tumor network. (d) Smoothing of the data
in c, analogous to b, yielding an estimate for locally averaged
MVD. Value ranges are 0-45 (black to white) for $\|\Delta P\|$ and
0-0.4 for the MVD. Note that in d the local MVD is highest (brightest)
in regions in the upper half of the network, and in b the local
precussure gradient is highest (brightest) also in regions in the
upper half of the network.
\bigskip

\textbf{\figinline{14}:}
Shows probability distributions in log-log plots, for
(a) the local MVD, given as the local average over 
250\micron wide boxes, (b) the volume of necrotic 
tissue clusters, defined as the number of sites 
in respective connected components of dead tissue.
(c) the volume of vessel hot-spot areas, defined
as the connected components of regions where the
local MVD exceeds a threshold (0.15). The distributions
are generated by binning observed values in histograms
over 40 simulation runs (at \timeis{1200}). Note that
the distributions show algebraic decay. In this instance
in particular with the same exponent within the error
bounds which are of the order of 2\%.
\bigskip

\textbf{\figinline{15}:}
Snapshots of the drug-flow simulation using the configuration A
tumor at \timeis{1000}. The color code shows the drug concentration as
indicated on the scale bar. Note that drug reaches most parts of the
vasculature quickly in a few seconds at the maximum concentration. In
the T=6.6s snapshots a few vessels in the boundary region are not yet
perfused. The time until drug saturates the complete vasculature is of
the order of minutes. We can observe this behavior generally for all
resulting networks.
\bigskip

\textbf{\figinline{16}:}
(a) shows the amount
of vessels given as the fraction of total tumor network 
length that has been exposed to a drug concentration
larger than the indicated concentration $c_{max}$ (at t=20s of
the drug simulation). The black line represents the average
over 40 runs with different networks.
(b) shows the 
amount of vessels given as the fraction of total tumor 
network length that has been exposed to a drug concentration
larger than 0.25 for longer than indicated time $t_e$.
The black line represents the average over 40 runs with 
different networks. Like in \figinline{15}, tumor 
networks from the base case at \timeis{1000} where used.
\bigskip

\newpage
\centering
\vfil
\includegraphics[width=5cm]{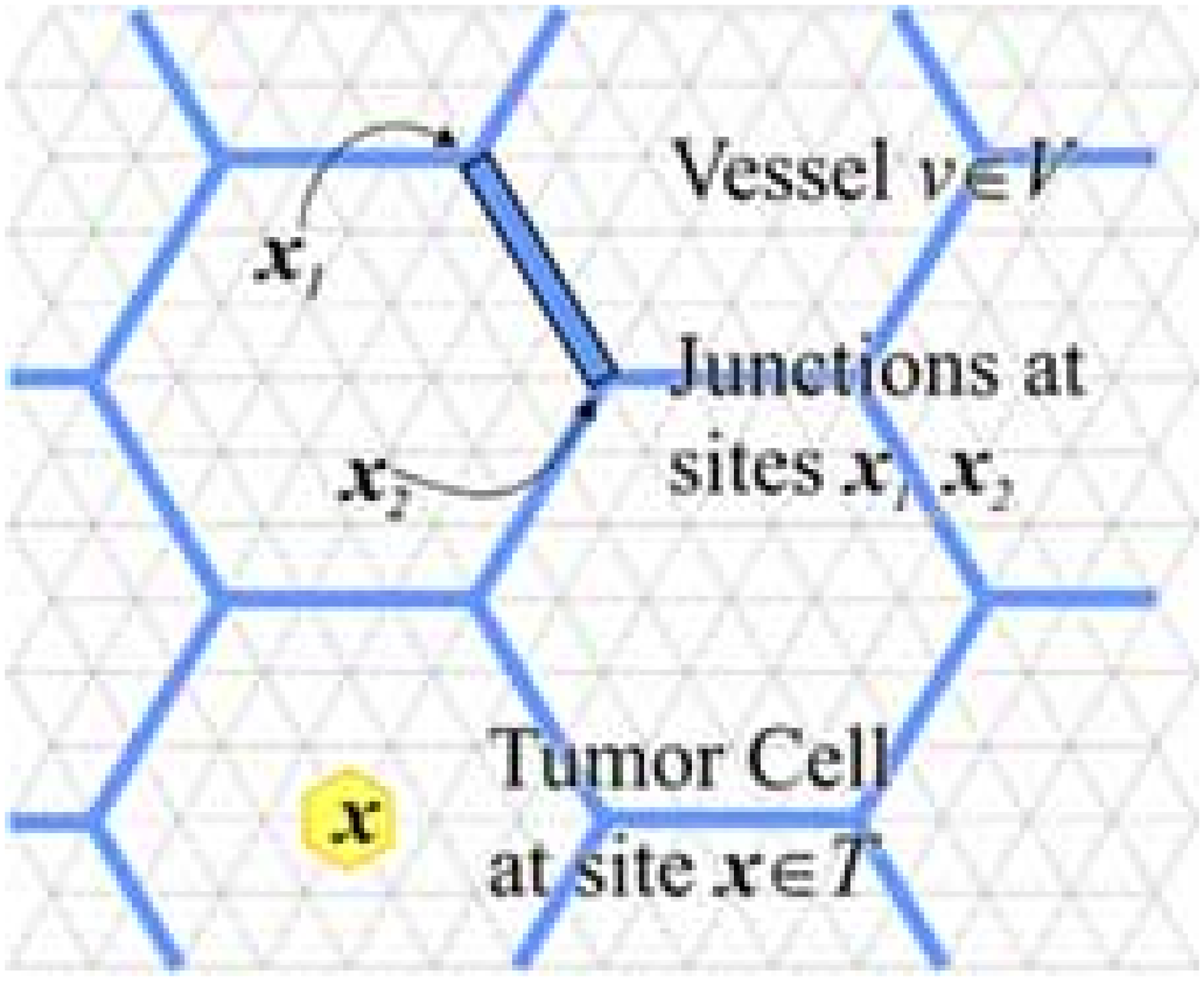}
\vfil
{\huge \bf Fig. 1}

\newpage
\centering
\vfil
\includegraphics[width=12cm]{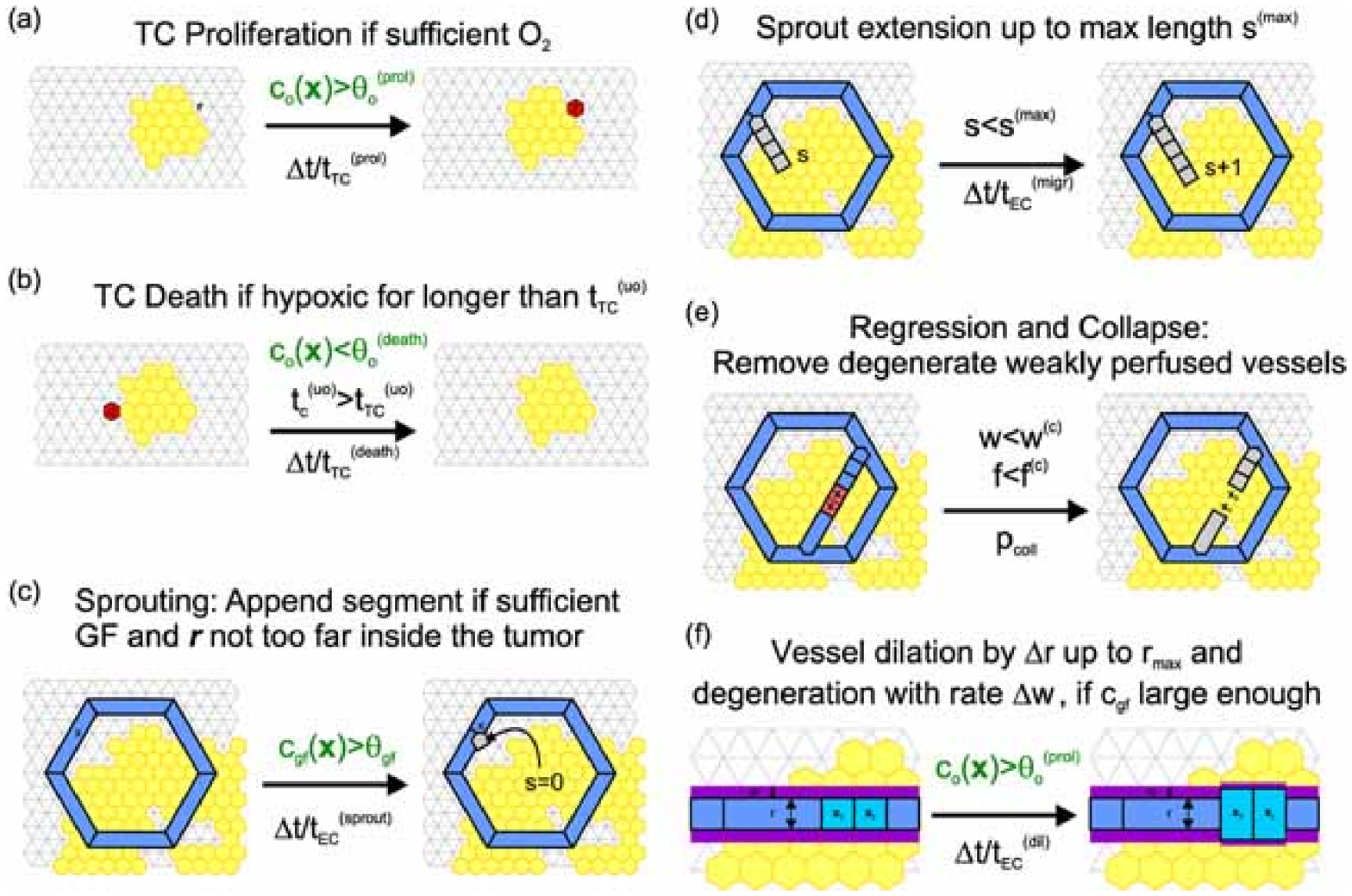}
\vfil
{\huge \bf Fig. 2}

\newpage
\centering
\vfil
\includegraphics[width=16cm]{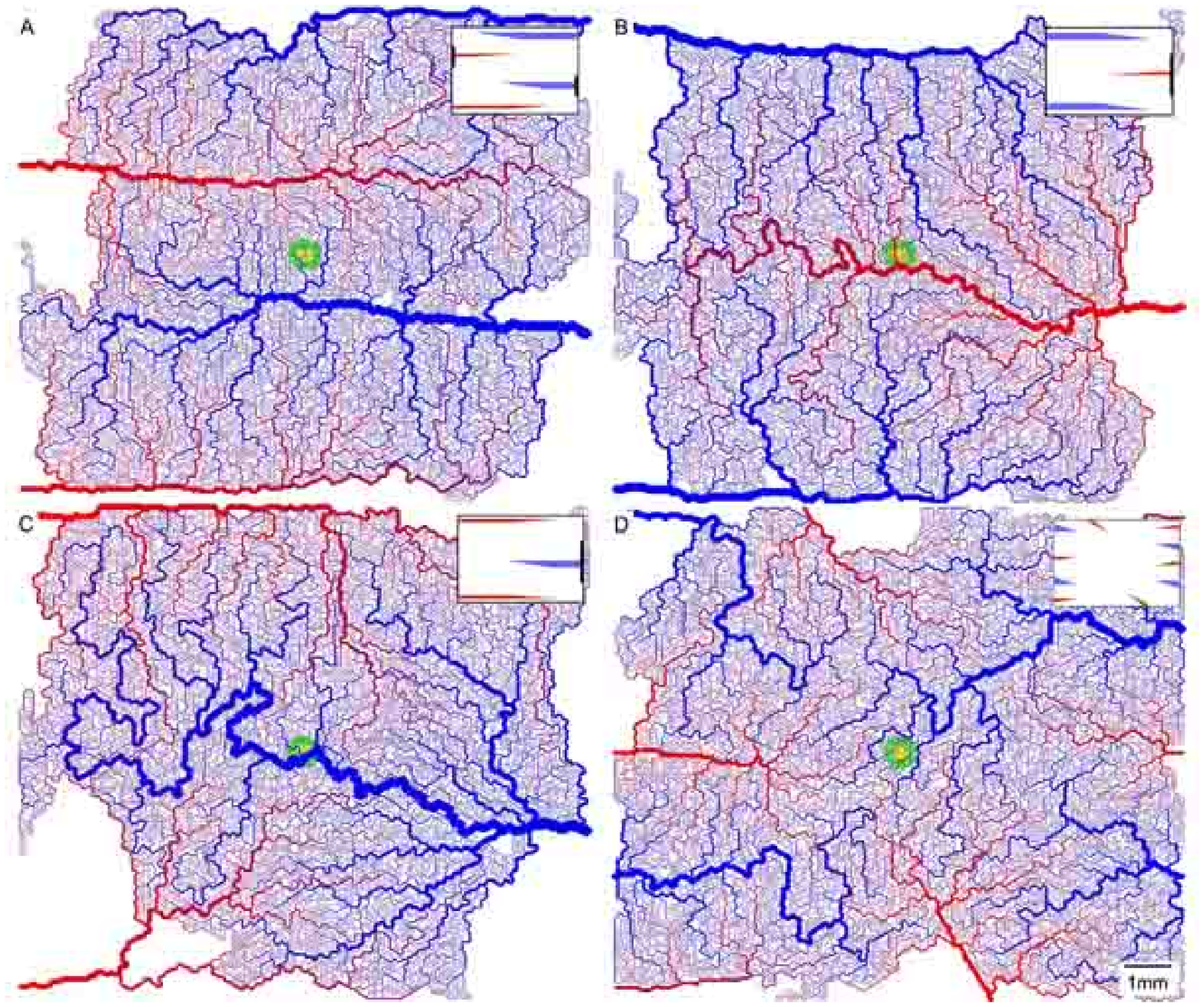}
\vfil
{\huge \bf Fig. 3}

\newpage
\centering
\vfil
\includegraphics[width=16cm]{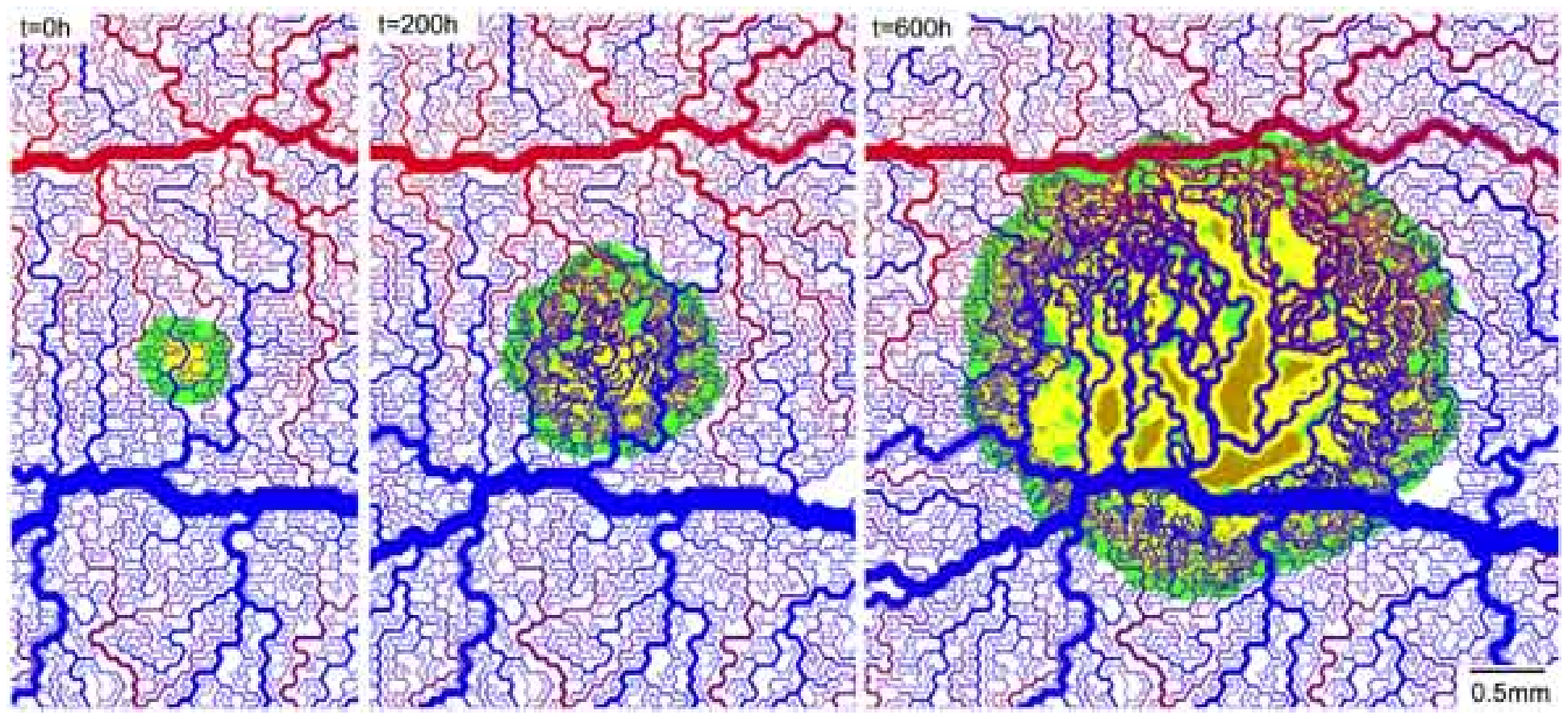}
\vfil
{\huge \bf Fig. 4}

\newpage
\centering
\vfil
\includegraphics[width=16cm]{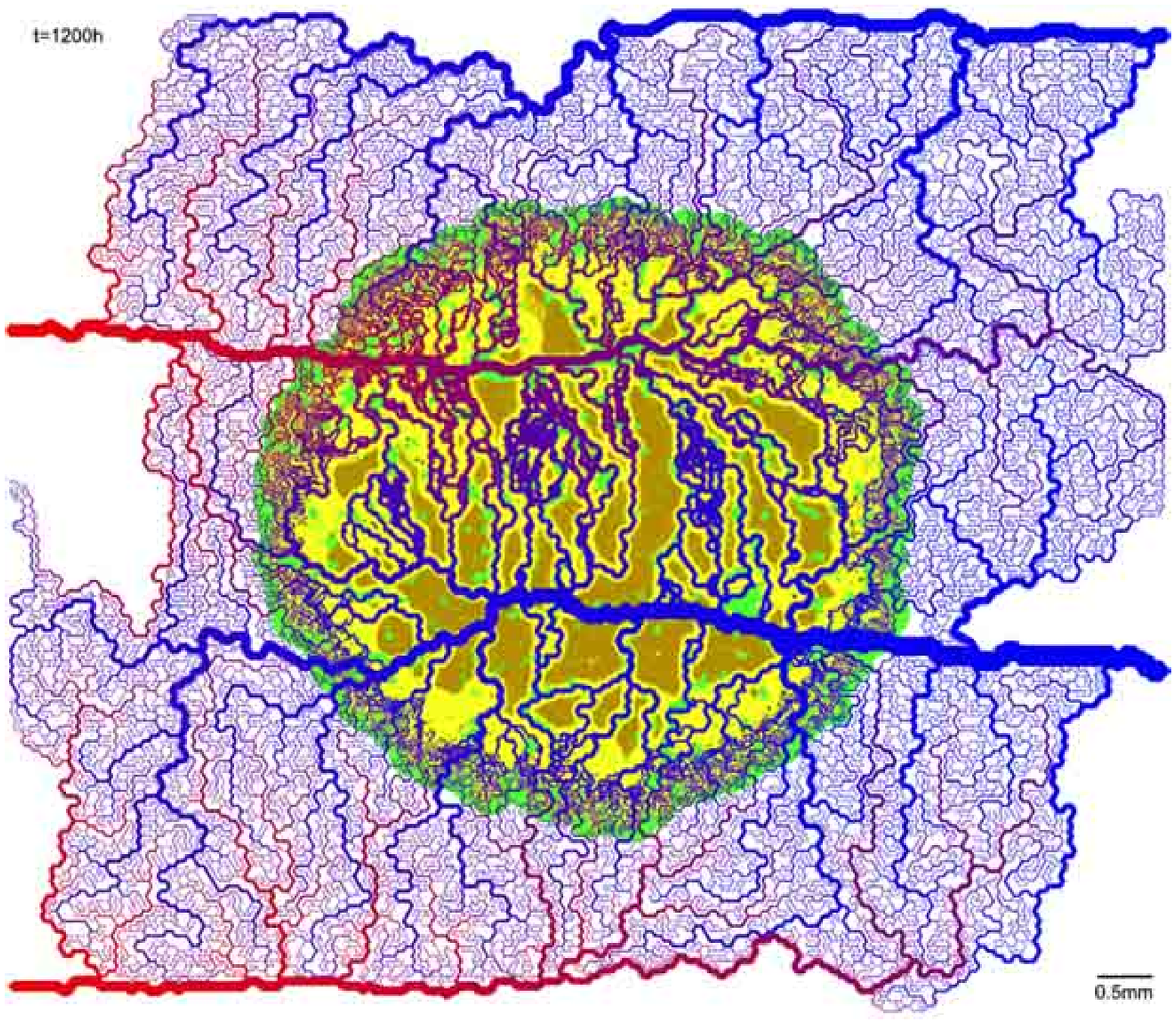}
\vfil
{\huge \bf Fig. 5}

\newpage
\centering
\vfil
\includegraphics[width=16cm]{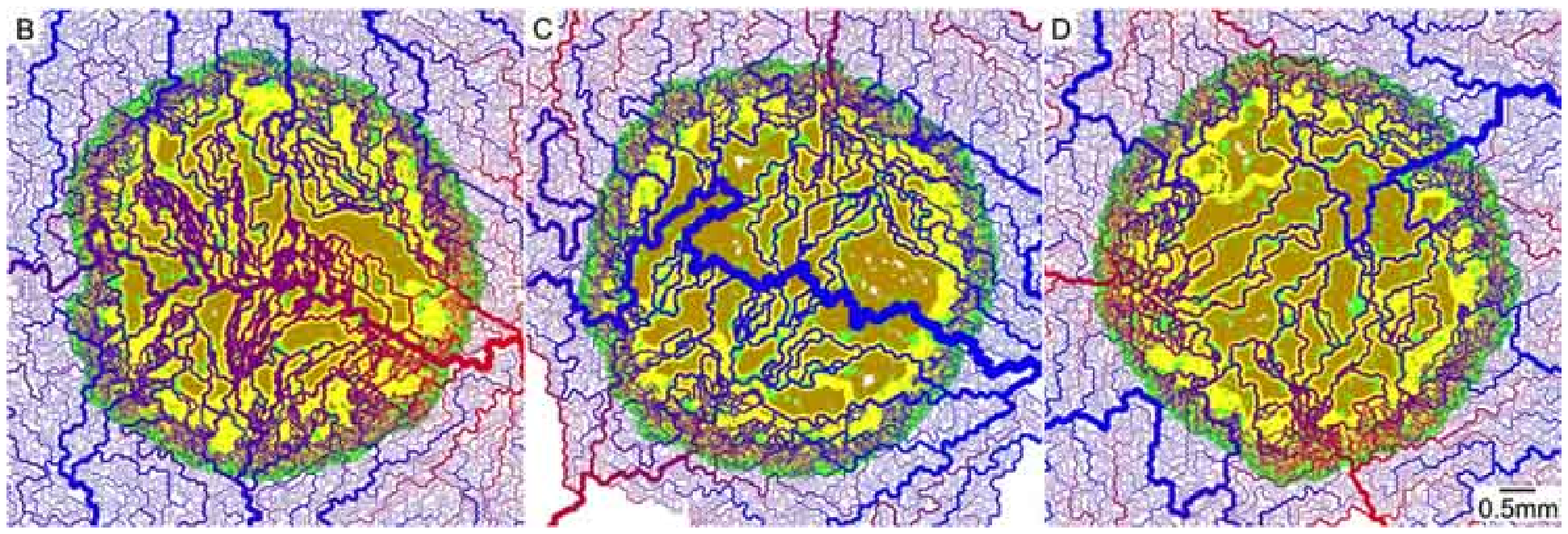}
\vfil
{\huge \bf Fig. 6}

\newpage
\centering
\vfil
\includegraphics[width=6cm]{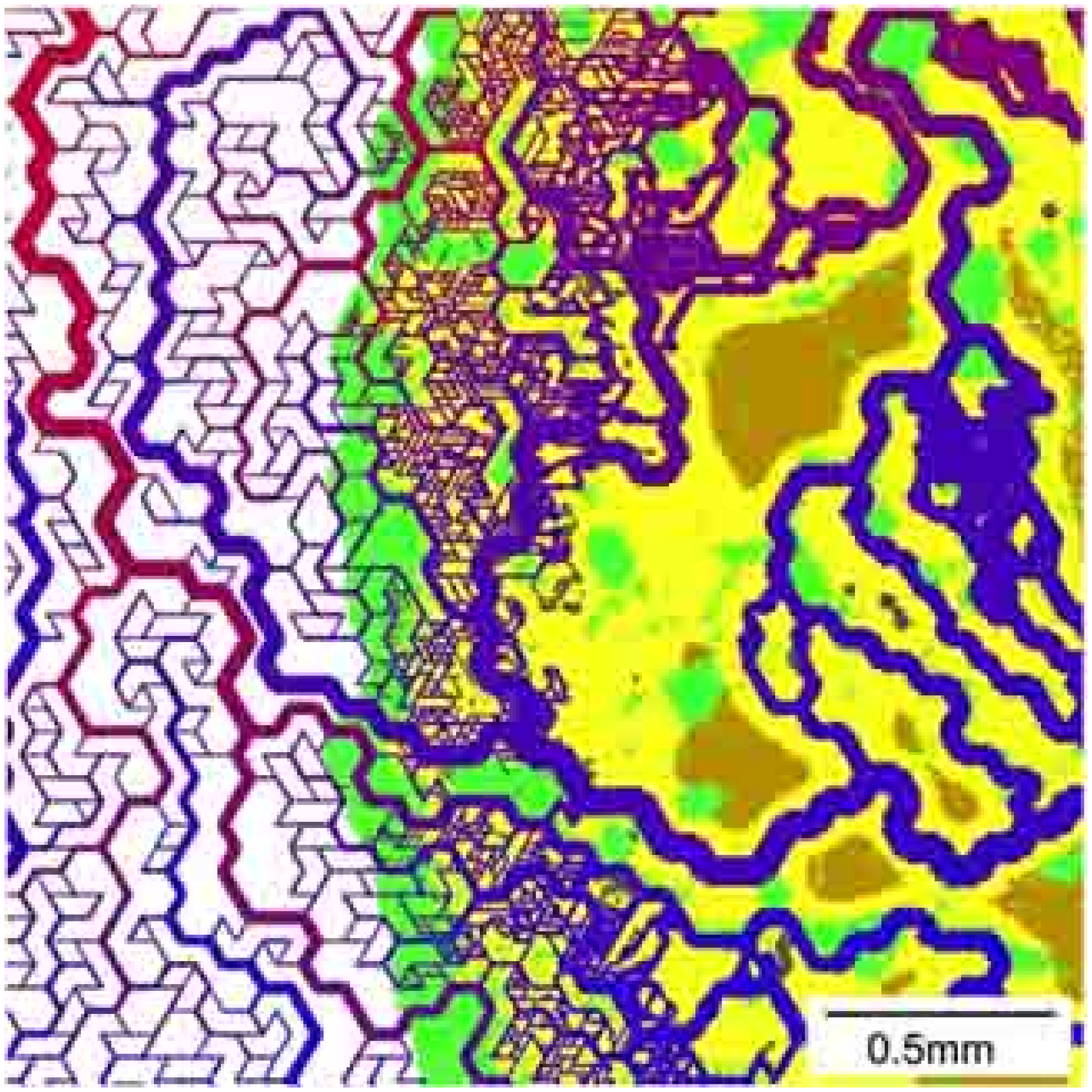}
\vfil
{\huge \bf Fig. 7}

\newpage
\centering
\vfil
\includegraphics[width=8cm]{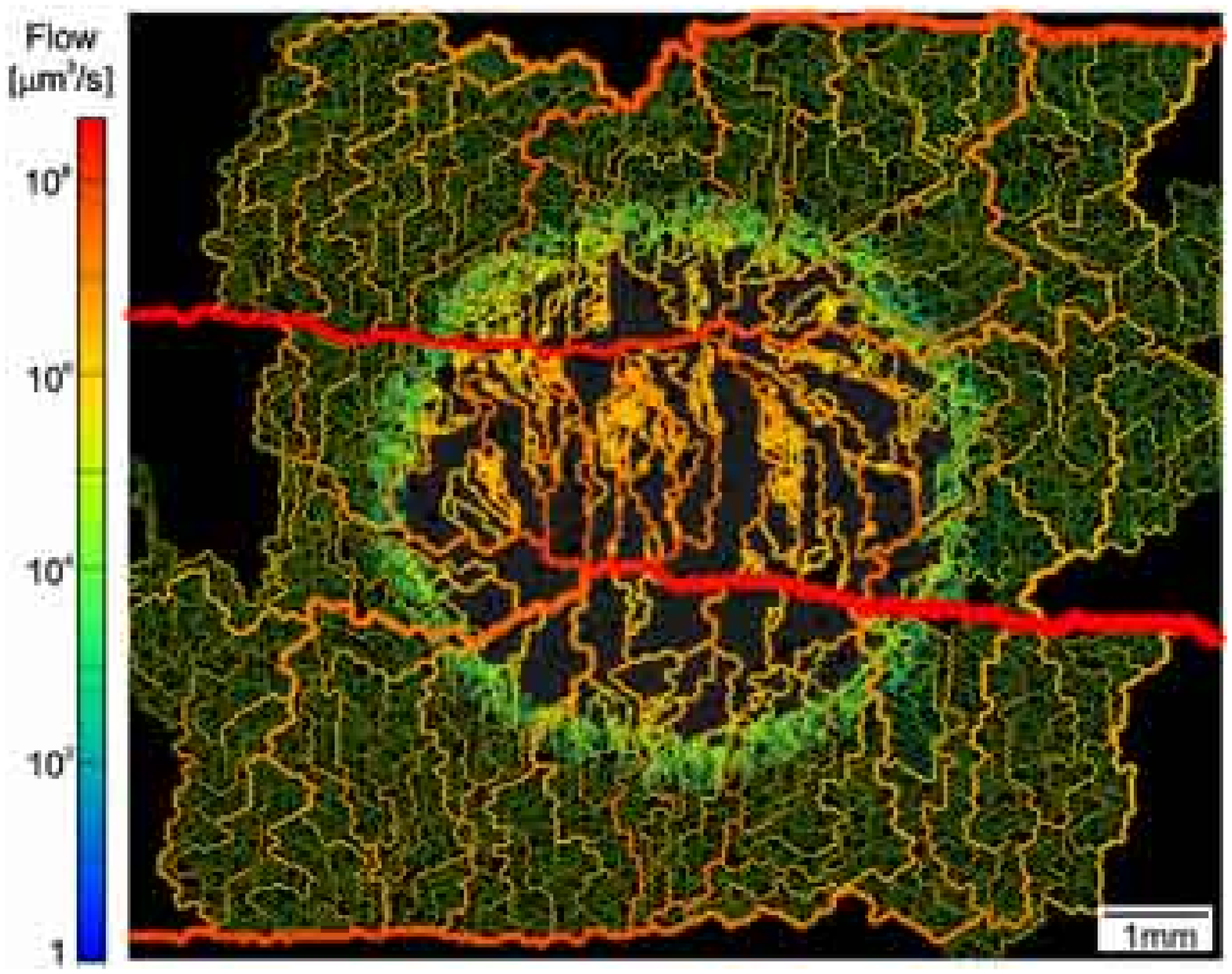}
\vfil
{\huge \bf Fig. 8}

\newpage
\centering
\vfil
\includegraphics[width=16cm]{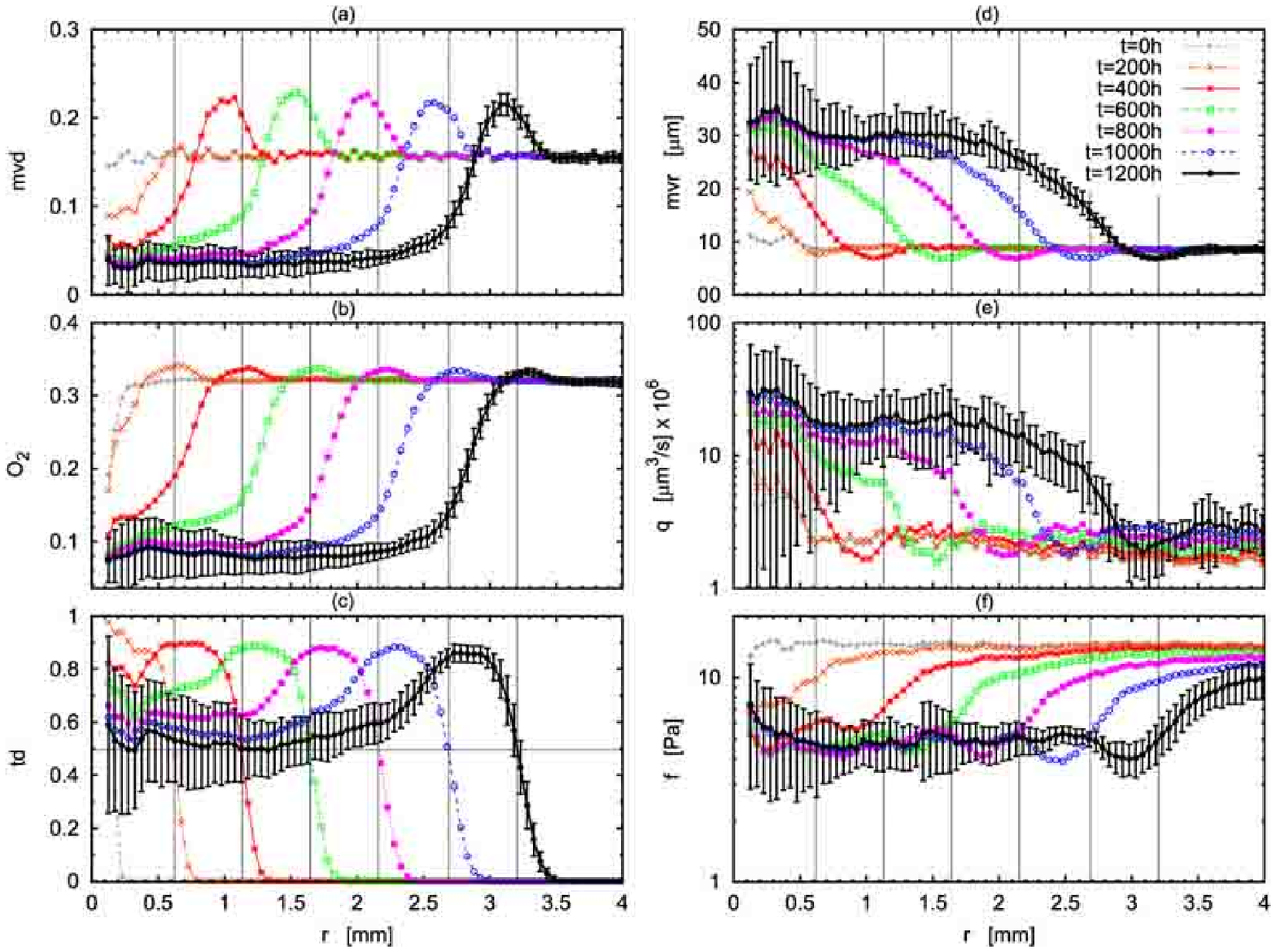}
\vfil
{\huge \bf Fig. 9}

\newpage
\centering
\vfil
\includegraphics[width=10cm]{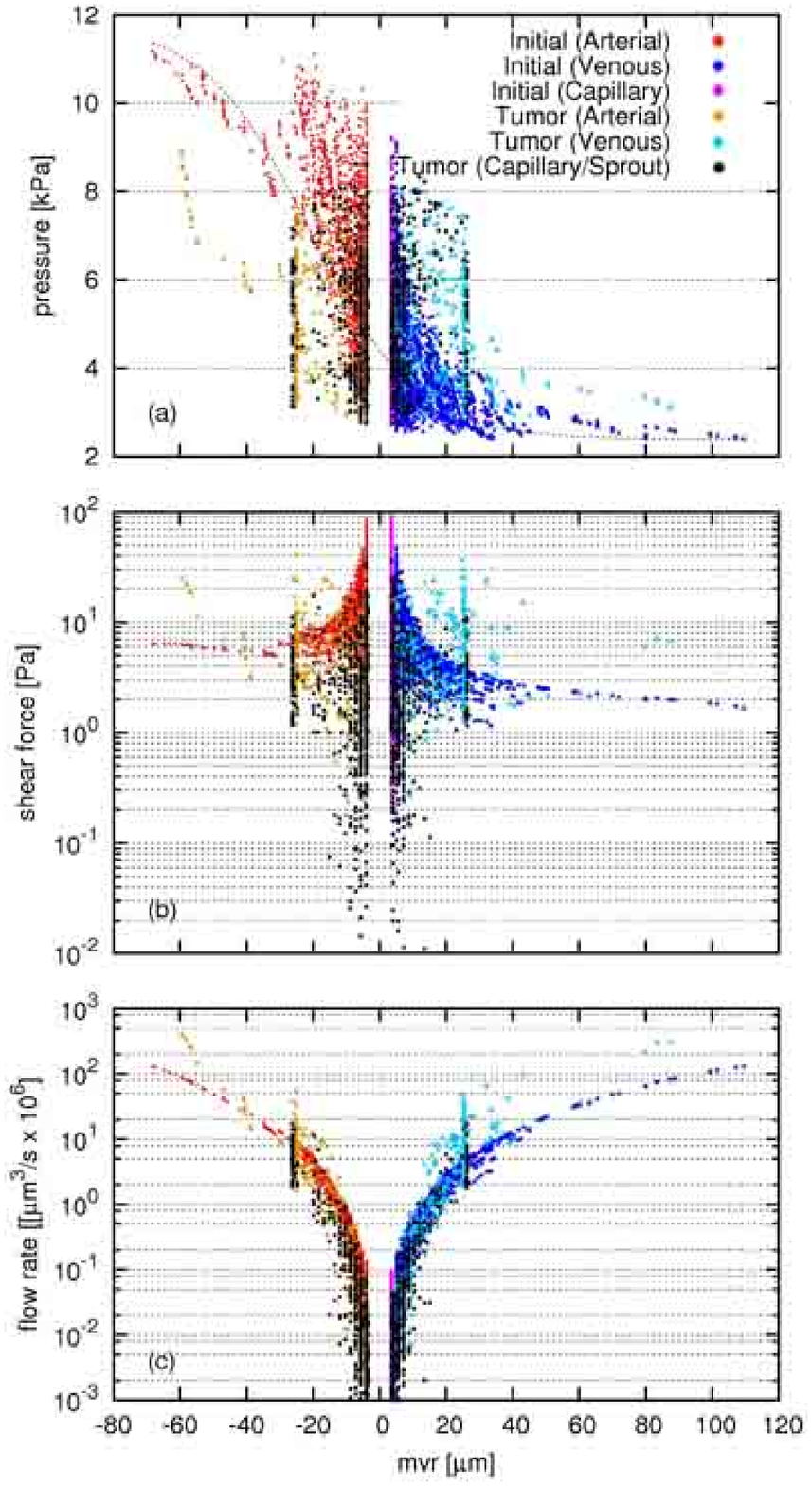}
\vfil
{\huge \bf Fig. 10}

\newpage
\centering
\vfil
\includegraphics[width=6cm]{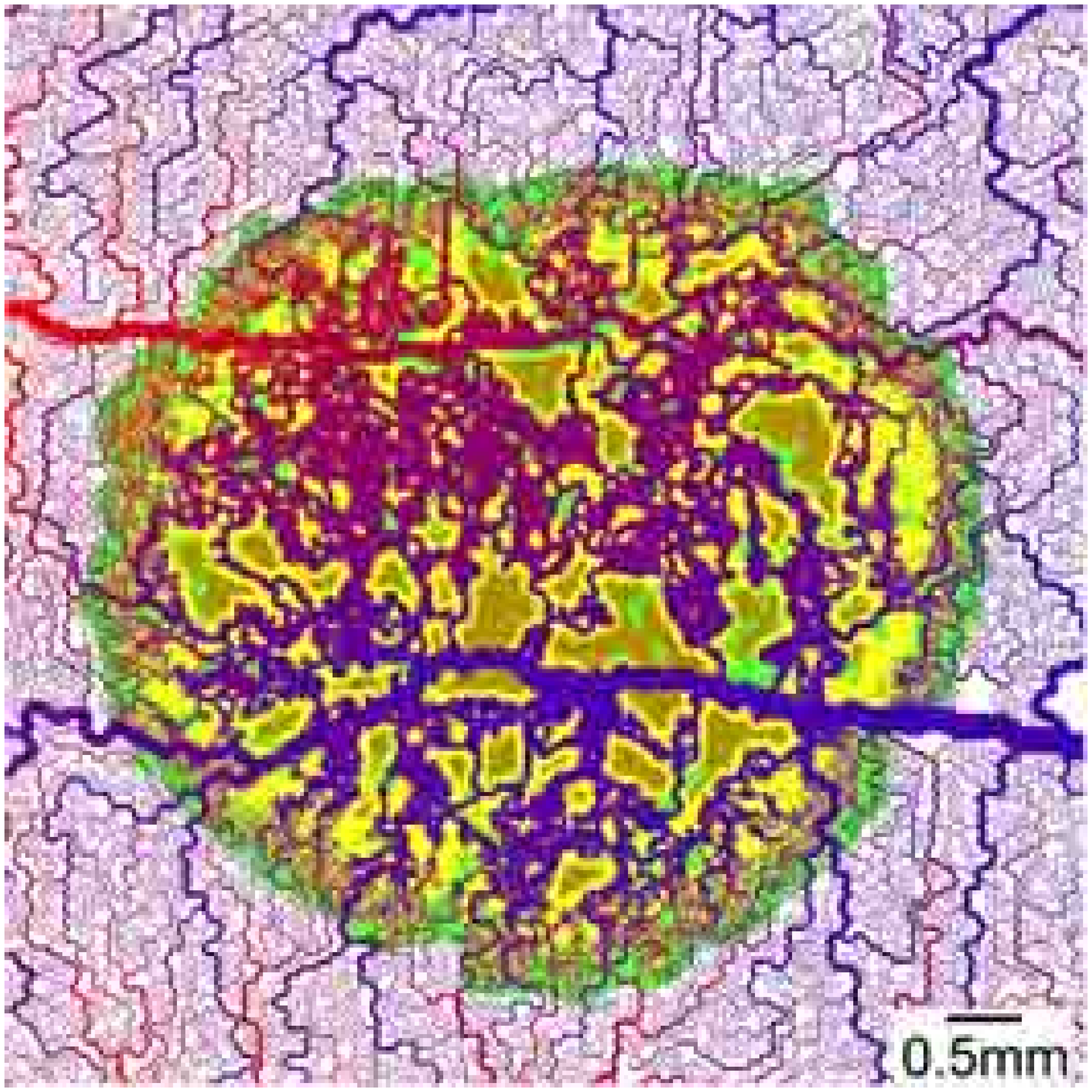}
\vfil
{\huge \bf Fig. 11}

\newpage
\centering
\vfil
\includegraphics[width=10cm]{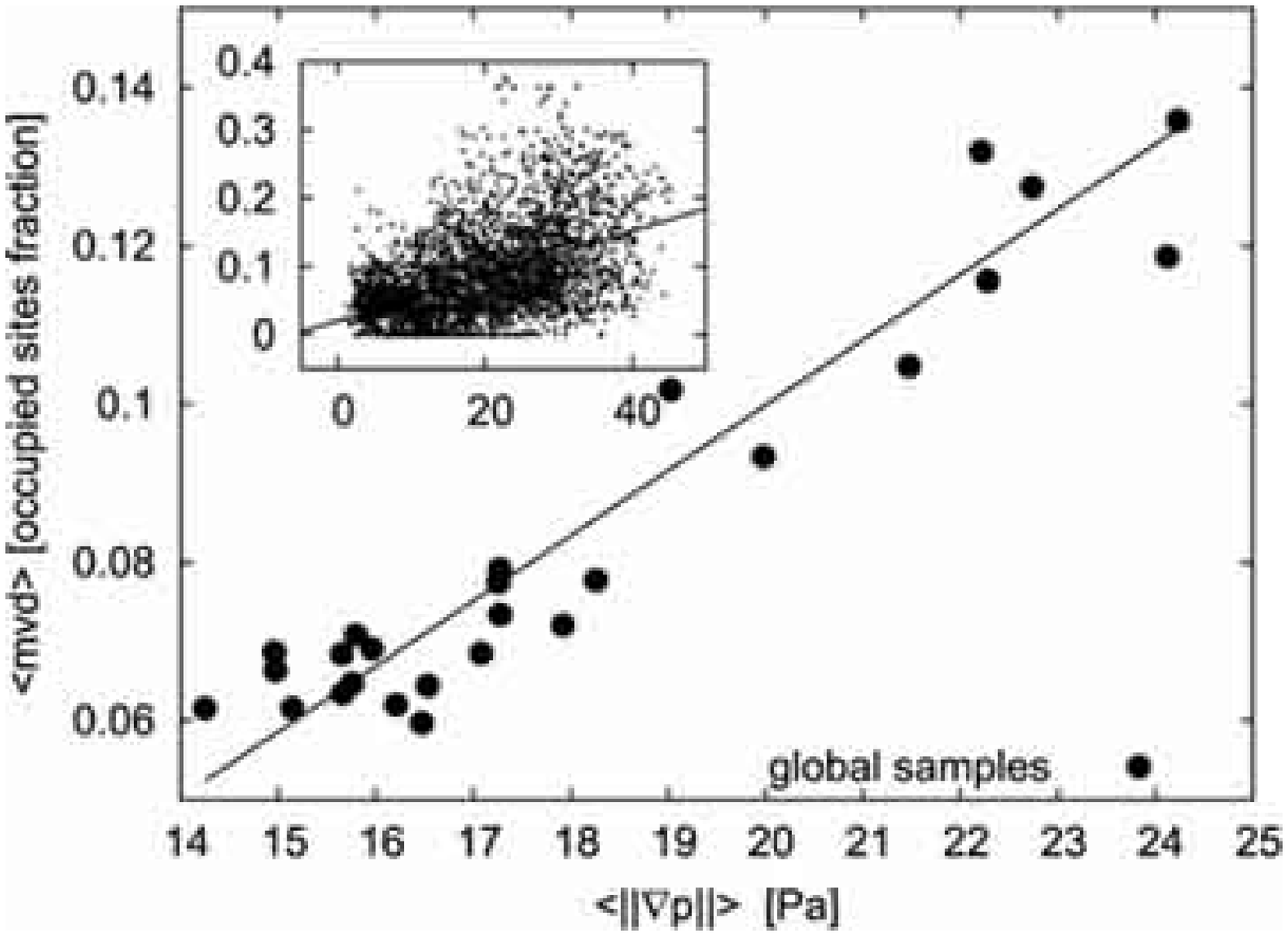}
\vfil
{\huge \bf Fig. 12}

\newpage
\centering
\vfil
\includegraphics[width=8cm]{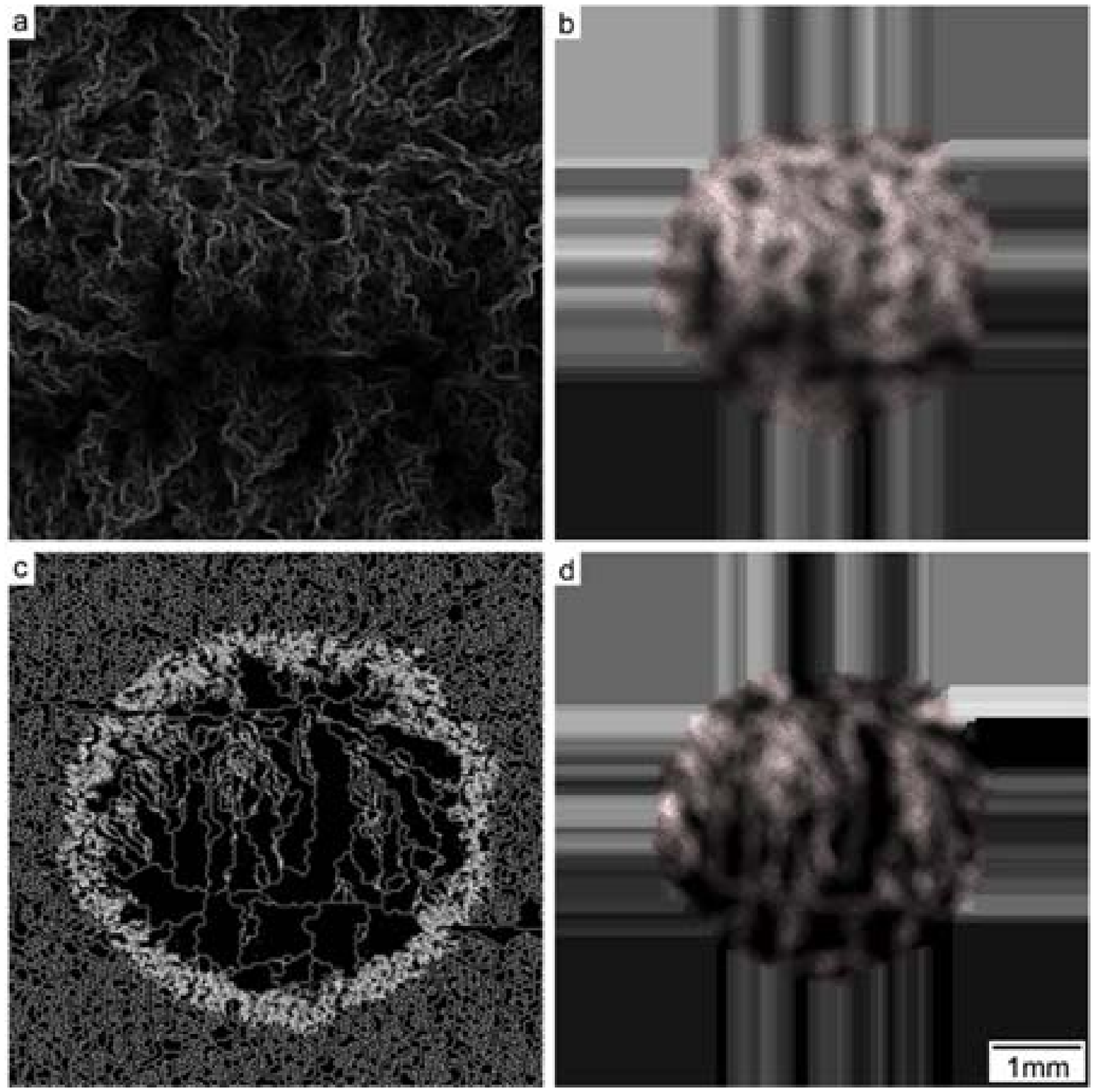}
\vfil
{\huge \bf Fig. 13}

\newpage
\centering
\vfil
\includegraphics[width=16cm]{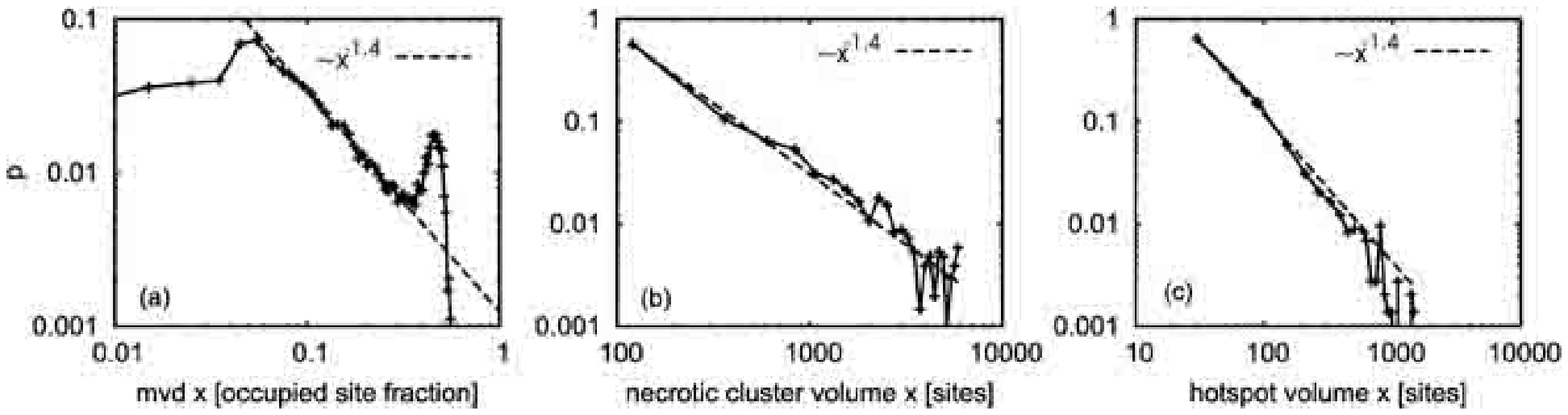}
\vfil
{\huge \bf Fig. 14}

\newpage
\centering
\vfil
\includegraphics[width=16cm]{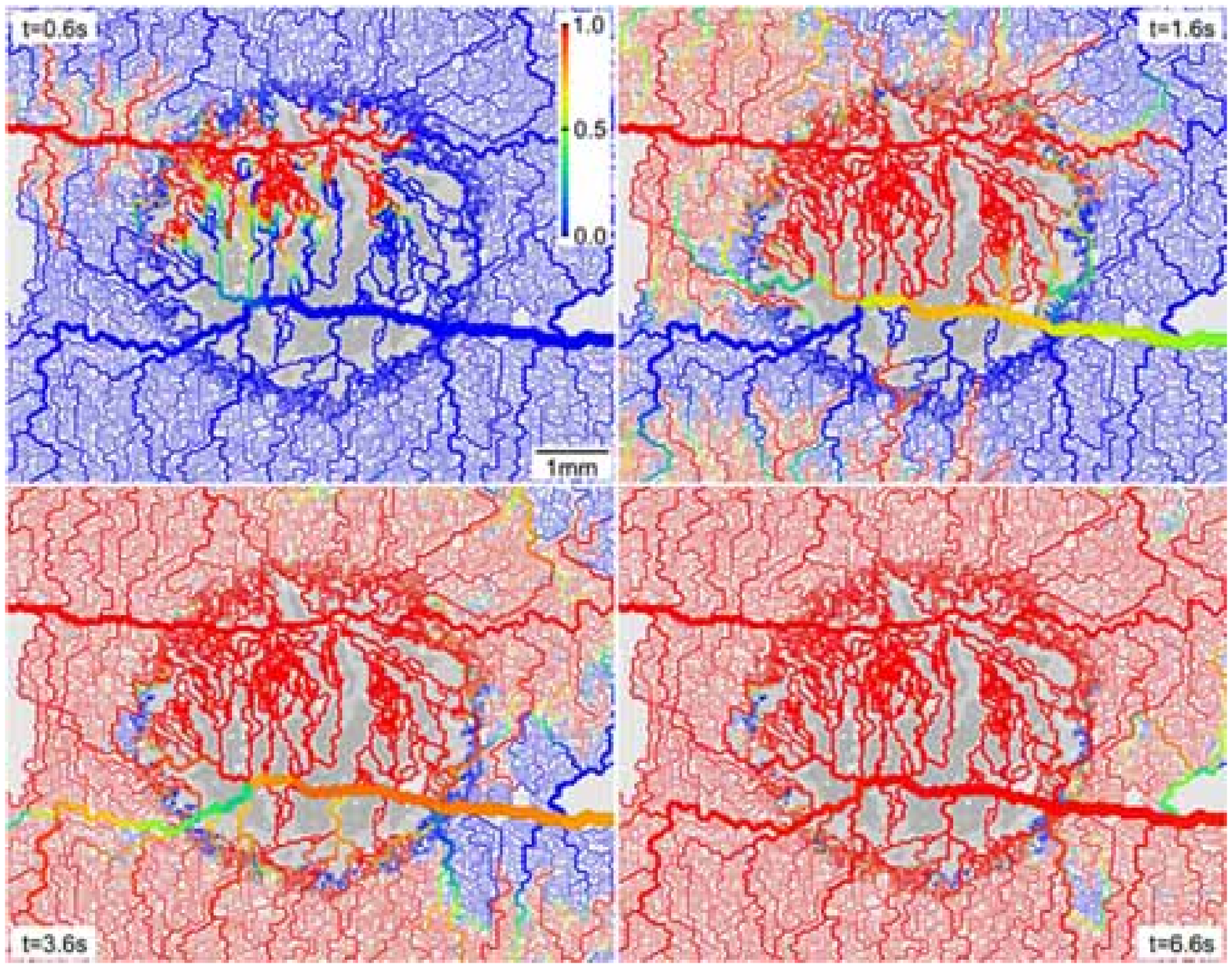}
\vfil
{\huge \bf Fig. 15}

\newpage
\centering
\vfil
\includegraphics[width=8cm]{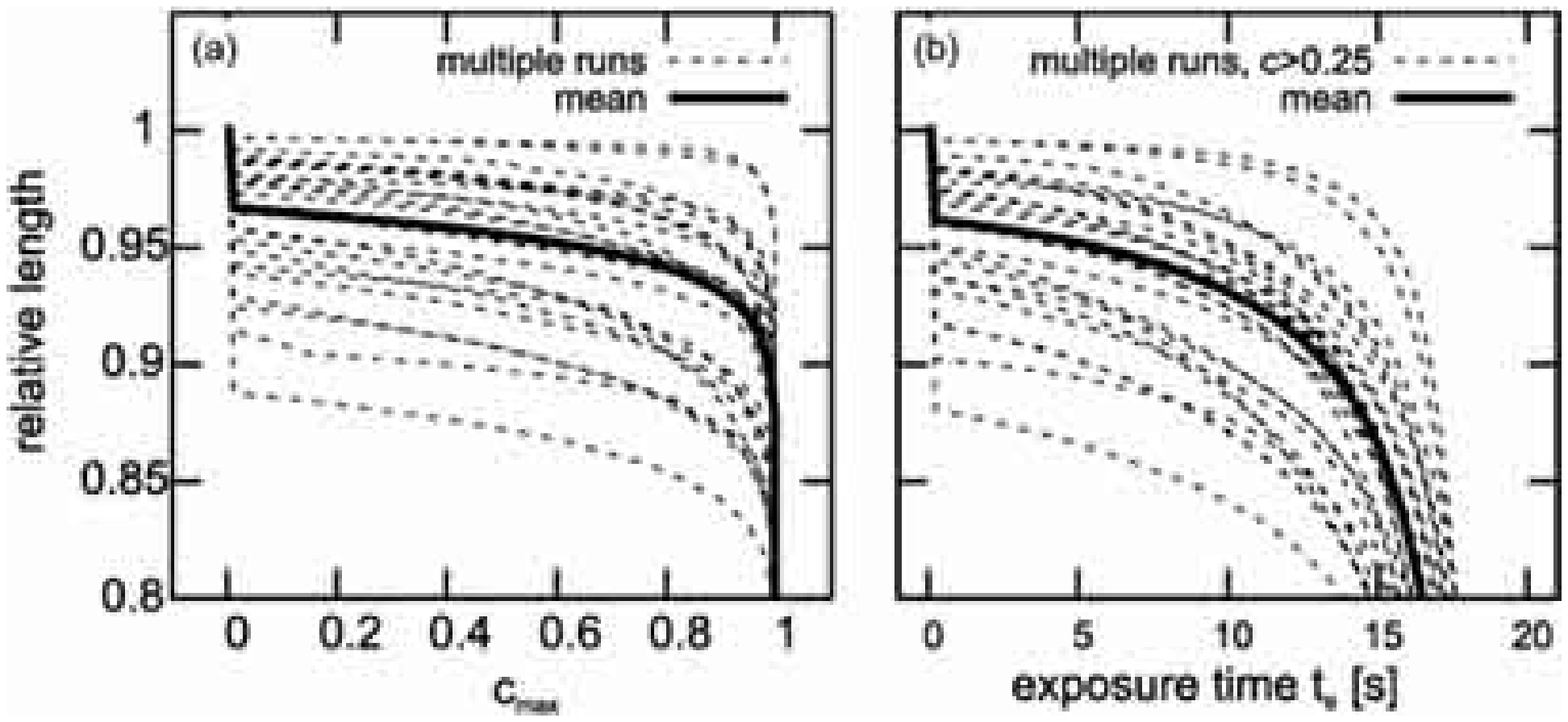}
\vfil
{\huge \bf Fig. 16}

\end{document}